\documentclass[letterpaper,11pt]{article}
\pdfoutput=1

\usepackage{amsmath,amssymb}
\usepackage{epsfig}
\usepackage{cite}

\def\Dslash{\hspace{3pt}\raisebox{1pt}{$\slash$} \hspace{-9pt} D}

\begin{document}

\baselineskip=18pt


\thispagestyle{empty}
\vspace{20pt}
\font\cmss=cmss10 \font\cmsss=cmss10 at 7pt


\hfill
\vspace{20pt}

\begin{center}
{\Large \textbf
{The Discrete Composite Higgs Model}}
\end{center}

\vspace{15pt}
\begin{center}
{\large Giuliano Panico and Andrea Wulzer} \vspace{20pt}

\textit{Institute for Theoretical Physics, ETH, CH-8093, Zurich, Switzerland}

\end{center}

\vspace{20pt}
\begin{center}
\textbf{Abstract}
\end{center}
\vspace{5pt} {\small
We describe a concrete, predictive incarnation of the general paradigm of a
composite Higgs boson, which provides a valid alternative to the standard
holographic models in five space-time dimensions. Differently from the latter,
our model is four-dimensional and simple enough to be implemented in an
event generator for collider studies. The model is inspired by dimensional
deconstruction and hence it retains useful features of the five-dimensional
scenario, in particular, the Higgs potential is finite and calculable. Therefore our setup,
in spite of being simple, provides a complete description of the composite
Higgs physics. After constructing the model we present a first analysis of its
phenomenology, focusing on the structure of the Higgs potential, on the
constraints from the EWPT and on the spectrum of the new particles.
}

\vfill\eject
\noindent


\section{Introduction}

The Hierarchy Problem, in combination with the LEP data on precision ElectroWeak (EW) physics,
leads rather naturally to the idea of a composite Higgs boson. This paradigm, first
proposed in \cite{gk}, has received considerable attention in the recent years
\cite{CHM1,CHM2,CHM2flat,Contino:2006nn,Giudice:2007fh,Barbieri:2007bh}.
The idea, in very broad terms, is the following:
the Higgs boson emerges as the bound state of a new strongly-interacting sector,
in particular it is the (pseudo) \emph{Nambu-Goldstone Boson} (pNGB) associated with
the spontaneous breaking
of the global symmetry group $G$ of the new sector. In this construction, differently from
technicolor, the SM group is not broken by the strong dynamics. The unbroken subgroup
$H$ of $G$ therefore contains the SM ${\textrm{SU}}(2)_L\times {\textrm{U}}(1)_Y$
and the coset space $G/H$ is chosen such as to include at least one
doublet of Higgs scalars. It is the VEV of this field that breaks spontaneously, as
in the ordinary Higgs model, the SM group. The ``minimal'' $G/H$ coset with these features,
which also contains a custodial ${\textrm{SO}}(4)$ symmetry that keeps $\delta\rho$
under control, is ${\textrm{SO}}(5)/{\textrm{SO}}(4)$. We will restrict to this minimal choice
in the following.

While the general idea is quite simple and clear, it is rather complicated to formulate
explicit and predictive incarnations of the composite Higgs scenario. The problem,
obviously, is modeling the dynamics of the strongly-interacting sector, a task for
which no definite recipe exists. We have however robust expectations on how this
composite Higgs model should look like \cite{Giudice:2007fh}, based first of all on
symmetry and large-$N_c$ considerations, but also on the experience with QCD.
We expect that the strong sector delivers, on top of the Goldstone boson Higgs,
resonances of typical mass $m_\rho$ which interact among each other and with
the Higgs with a ``typical coupling'' $g_\rho$. In spite of emerging from a strongly-interacting
dynamics, this coupling is not necessarily large and close to the perturbative bound of
$4\pi$. Provided we assume that the underlying dynamics of our strong sector is a
large-$N_c$ gauge theory, indeed,  $g_\rho$ can be made parametrically small as
it scales like $g_\rho\simeq4\pi/\sqrt{N_c}$. Whether $g_\rho$ is large or small becomes at
this point a phenomenological question. The answer given in
ref.~\cite{CHM1,Giudice:2007fh}
is that $g_\rho$ is preferentially large because this helps the model to pass the EW
precisions tests (EWPT).
We can restrict, for definiteness, the $g_\rho$ coupling in the range $1<g_\rho<4\pi$.

The above general considerations are already sufficient to develop a comprehensive but
\emph{qualitative} description of the composite Higgs model, the so-called ``SILH''
approach \cite{Giudice:2007fh}. \footnote{Within the SILH (see also \cite{Barbieri:2007bh})
quantitative predictions are possible but only for those properties that follow directly from
the Goldstone boson nature of the Higgs, {\it{i.e.}}~from the structure of the non-linear
$\sigma$-model. In this framework there is no quantitative description of the resonances, which is what
is mostly needed for the LHC phenomenology.}
To be quantitative, a specific and explicit construction is needed. At present, the only
candidates are the ``holographic'' models, formulated as gauge theories in five space-time
dimensions (5d), which have been extensively explored \cite{CHM1,CHM2,CHM2flat}.
The aim of the present paper is to describe an alternative and \emph{simpler} possibility.
The need of simplifying the 5d models has already been recognized in the literature
\cite{Contino:2006nn}: their objective technical complication makes them very difficult
to implement in an event generator, and therefore not really suited for the LHC collider
phenomenology. This technical problem goes together with a conceptual limitation. The
5d models are unnecessarily complicated because they contain much more information
than what could be tested at the LHC. Namely, they incorporate Kaluza-Klein (KK) towers of
resonances, with increasing mass, and formally describe the dynamics of each of these
particles. Only the very first few resonances, however, are accessible at the LHC, retaining
also the heavier ones is a redundancy of the 5d construction that we aim to overcome.

Our approach is inspired by the 5d models and by the idea of \emph{dimensional deconstruction}
\cite{ArkaniHamed:2001ca},
which consists in discretizing the extra coordinate in a finite number of points, or \emph{sites}.
We therefore denote our setup as the ``Discrete'' Composite Higgs Model (DCHM) and, depending
on the number of deconstruction sites, we will be talking about the two-site or the three-site models.
The DCHM is a completely four-dimensional (4d) theory, and each site is associated to a
set of 4d degrees of freedom which correspond, roughly speaking,
to one level of the KK tower. The DCHM provides a simplified version of the 5d model
where only few KK levels, {\it i.e.}~few resonances of the strong sector, are included.
Given that the first level, the ``zero-modes'', corresponds to the ordinary SM particles, the
two-site model describes only \emph{one} set of strong sector resonances; \emph{two} replicas
of the latter appear in the case of three sites and so on. \footnote{Even though this
simplified discussion might suggest the contrary, the deconstructed model is deeply different
from a naive truncation of the KK tower, which is the approach adopted in ref.~\cite{Contino:2006nn}.
In the Conclusions we will comment more on the relation with our construction, we can however
anticipate that the difference lies in the peculiar symmetry structure of the deconstructed model,
which is badly broken by the truncation.} An approach similar to ours has already been
applied to the 5d Higgsless models (see for instance \cite{Chivukula:2006cg}) and in the
context of Little Higgs theories \cite{Cheng:2006ht,Foadi:2010bu,Baumgart:2007js}. \footnote{Further comments on the relation of our
model with the ones of ref.~\cite{Cheng:2006ht,Foadi:2010bu} are postponed to the Conclusions.}

Aiming to simplicity, as we do in the present paper, the two-site model seems the more convenient
choice because already the second level of resonances is probably rather heavy
and not easy to observe at the LHC. However, we want to build a model which is not only
simple, but also \emph{predictive}, or better as predictive as the 5d ones for what the EW and
the Higgs physics is concerned. In particular, the relevant observables that we would like
to be calculable in our model, like they are in the 5d ones, are the \emph{potential} of the
composite Higgs boson and the ${\widehat S}$ and ${\widehat T}$ parameters of EWPT
\cite{Peskin:1991sw}. If this was the case, we would be able to incorporate rigorously, and not
only to estimate, the constraint on the parameter space that follows from the EWPT
and also from the direct measurement of the Higgs mass, when it will become available.
In case of discovery, one could imagine \emph{fitting} the masses and couplings of the
new resonances together with the Higgs mass and, if measured, with other parameters
of the potential such as the trilinear Higgs coupling. It turns out from the analysis that we will
present in section~\ref{th} that in the two-site model the ${\widehat S}$ and ${\widehat T}$
parameters are indeed calculable, but not the Higgs potential. The best
compromise among simplicity and predictability will therefore turn out not to be the two-site
but the \emph{three-site} DCHM, in which also the potential becomes calculable.

The paper is organized as follows. In section~\ref{th} we construct the DCHM and discuss
in great detail its symmetry structure and its interpretation as a calculable effective field theory.
The main practical outcome will be, as previously mentioned, to establish the calculability
of the ${\widehat S}$ and ${\widehat T}$ parameters and of the Higgs potential.
Section~\ref{th} is rather technical, the reader interested in phenomenology could skip
it in a first reading and pass directly to section~\ref{3S}, where a self-contained phenomenological
study of the three-site DCHM is presented. In particular, we discuss the properties of the
Higgs potential, the constraints coming from the EWPT and the general features of the spectrum
of the strong sector resonances. We discuss our conclusions in section~\ref{conc}
and in Appendix~\ref{gener} we summarize the conventions adopted for the ${\textrm{SO}}(5)$ generators.
Appendix~\ref{NDA} contains an alternative derivation of the Naive
Dimensional Analysis (NDA) formula \cite{Manohar:1983md}, which we employ extensively in section~\ref{th}.

\section{Models of composite Higgs}\label{th}

The best way to introduce our model is to proceed in a constructive manner, starting from
the simplest possible description of the composite Higgs scenario and progressively adding the
necessary ingredients. This leads to a simpler exposition, in which the required technical tools
are introduced gradually, and moreover it allows us to clarify the relation of our DCHM with
other 4d approaches \cite{Barbieri:2007bh,Anastasiou:2009rv}.

\subsection{The non-linear $\sigma$-model}\label{sigmam}

The starting point is the non-linear $\sigma$-model of the
${\textrm{SO}}(5)/{\textrm{SO}}(4)$ coset, which obviously constitutes the ``minimal''
description of a composite Higgs. This model does not contain strong sector
resonances, but only the pNGB Higgs. To construct the $\sigma$-model we introduce,
as usual, a $\Sigma$-field
\begin{equation}
\displaystyle
\Sigma_I\,=\,U_{I5}\,,\;\;\;\;\;\;
U\,=\,
\exp\left[i\frac{\sqrt{2}}{f}\,\Pi_{\widehat{a}} T^{\widehat{a}}\right]
\,,
\label{sigmm0}
\end{equation}
which transforms in the fundamental representation of $\textrm{SO}(5)$. The
fields $\Pi_{\widehat{a}}$ are the four NGB Higgs fields, corresponding to the
four components of one Higgs doublet, and transform in the ${\mathbf 4}$ of
$\textrm{SO}(4)$. We will denote collectively as $T^A=\left\{{T}^a,\,\,T^{\widehat{a}}\right\}$
the unbroken (${T}^a\in {\textrm{Lie}}[H]$) and broken ($T^{\widehat{a}}$)
generators of $\textrm{SO}(5)$ in the fundamental representation,
normalized to $\textrm{Tr}\left[T^A T^B\right]=\delta^{AB}$.
The explicit form of the
generators is reported in Appendix~\ref{gener}. At the $2$-derivative order, the only
$\textrm{SO}(5)$-invariant term that can appear in the Lagrangian is
\begin{equation}
\displaystyle
{\mathcal L}^{\pi}\,=\,\frac{f^2}2\sum_{I}
\partial_\mu\Sigma_I\partial^\mu\Sigma_I\,.
\label{lagg0}
\end{equation}
where the normalization has been chosen such that the $\Pi_{\widehat{a}}$
kinetic term is canonical.

The coupling of the NGB Higgs to the SM gauge fields $W_\mu^{\alpha}$ ($\alpha=1,2,3$)
 and $B_\mu$ is introduced
by gauging the $\textrm{SU}(2)_L\times \textrm{U}(1)_Y$ subgroup of $\textrm{SO}(4)$
\footnote{We momentarily ignore the presence of the extra $\textrm{U}(1)_X$ charge in the definition
of the hypercharge and identify $Y=T_R^3$. The $\textrm{U}(1)_X$ does not play any role
until we introduce the SM fermions.}, {\it i.e.}~by substituting  the
ordinary derivative in eq.~(\ref{lagg0}) with the covariant one
\begin{equation}
D_\mu\Sigma \, =\, \partial_\mu\Sigma \,-\,i\,A_\mu \Sigma
\;\;\;\;\;\textrm{with}\;\;\;\;\; A_\mu\,=\,g\, W_\mu^\alpha T_L^\alpha
\,+\,g' B_\mu T_R^3 \,,
\label{cder}
\end{equation}
where the $\textrm{SU}(2)_L\times \textrm{U}(1)_Y$ generators $T_L^\alpha$ and
$T_R^3$ are defined in eq.~(\ref{eq:SO4_gen}).
One also has to include canonical kinetic (and interaction) terms for the gauge fields
\begin{equation}
\displaystyle
{\mathcal L}^{g}\,=\,-\frac{1}4
{\textrm{Tr}}\left[W_{\mu\nu}W^{\mu\nu}\right]
\,-\,\frac{1}4B_{\mu\nu}B^{\mu\nu}\,,
\label{lagg1}
\end{equation}
where we have defined $W_\mu=W_\mu^\alpha T_L^\alpha$ with
$W_{\mu\nu}=\partial_\mu W_\nu-\partial_\nu W_\mu - i g [W_\mu,W_\nu]$.

The Lagrangian ${\mathcal L}_0={\mathcal L}^{\pi}+{\mathcal L}^{g}$ we have constructed
is just the first of the infinite number of terms, ${\mathcal L}={\mathcal L}_0+\sum_i{\mathcal L}_i$,
that appear in our effective field theory. As a prerequisite for any quantitative statement on the effect
of these operators we need a power counting rule to estimate the size of their coefficients.
We assume the NDA power counting \cite{Manohar:1983md},
which assigns to each operator the size it receives from radiative corrections with a cut-off
$\Lambda$. The contribution from a diagram with $L$ loops takes the generic form (see Appendix~\ref{NDA})
\footnote{The only diagrams included in this estimate are those constructed with the
vertices of the ``leading order'' Lagrangian ${\mathcal L}_0$.}
\begin{equation}
\displaystyle
{\mathcal L}_i\,=\,\Lambda^2\,f^2\left(\frac{\Lambda}{4\pi f}\right)^{2L}
\left(\frac{\Pi}{f}\right)^{E_\pi}
\left(\frac{gW}{\Lambda}\right)^{E_W}
\left(\frac{\partial}{\Lambda}\right)^{d}\,
\left(\frac{g f}{\Lambda}\right)^{2\eta}\,,
\label{powcount}
\end{equation}
where $g$ and $W$ collectively denote, respectively, the $g$ and $g'$ couplings and the
gauge fields $W_\mu$ and $B_\mu$. The cutoff $\Lambda$ should
be set to its  NDA value $\Lambda=\Lambda_{\textrm{Max}}=4\pi f$ corresponding to
the scale at which the $\sigma$-model enters the strongly coupled regime.
We have however left $\Lambda$ as a free parameter because we want
to use the above equation not only to read the NDA
estimate of each given operator, but also its degree of divergence.

Another important
aspect of eq.~(\ref{powcount}) is the presence of the $(g f/\Lambda)^{2\eta}$ term, that by using
$\Lambda = 4\pi f$ can be rewritten as $(g^2/16\pi^2)^\eta$.
Each operator generically receives diagrammatic contributions of different
type, which lead to different values of $\eta$. Given that $g<4\pi$, the
leading contribution is the one with minimal value of $\eta$.
It is shown in  Appendix~\ref{NDA} that $\eta$ is necessarily
positive, $\eta\geq0$, but there might be obstructions in reaching the absolute minimum
$\eta=0$. In several cases that we will consider below, the symmetries enforce
$\eta\geq\eta_{\textrm{min}}>0$, which leads to a reduction of the degree of divergence
by the factor $(g f/\Lambda)^{2\eta_{\textrm{min}}}$ and to the corresponding
$(g^2/16\pi^2)^{\eta_{\textrm{min}}}$ suppression of the NDA estimate.

Several interesting aspects of the composite-Higgs scenario can be captured by the simple
model we have described (see \cite{Barbieri:2007bh}), but since we want to go beyond let us focus
on its limitations. First of all, it does not contain a description of the strong sector resonances,
which instead provide the key ingredient for the LHC phenomenology. Secondly, and related
to this, it does not allow to compute several important observables
such as the ${\widehat{S}}$ and ${\widehat{T}}$ EWPT parameter and
the Higgs mass $m_H$. It offers, in this sense, a limited predictive power.

Let us first discuss ${\widehat{S}}$, which, we remind, is proportional to the
derivative at zero momentum of the $W^3$-$B$ correlator. In our $\sigma$-model, this comes from
local operators with two derivatives and two gauge fields
such as
\begin{equation}
\displaystyle
\frac{c_S}{16\pi^2}{\mathcal O}_{S}\,=\,\frac{c_S}{16\pi^2}
\Sigma^tA_{\mu\nu}A^{\mu\nu}\Sigma\,\supset
\,
-\frac{1}{16\pi^2}
\frac{c_S}2\sin^2\left(\langle \Pi_4 \rangle/f\right)
g g'\,W^3_{\mu\nu} B^{\mu\nu}
\label{localS}
\end{equation}
where $\langle\Pi_4\rangle \simeq v = 246\,{\rm GeV}$ in the limit of small $v/f$.
The coefficient $c_S$, following the NDA estimate of eq.~(\ref{powcount}),
is a parameter of order unity and it has vanishing degree of divergence at one loop,
which means, of course, that it diverges logarithmically.
The contribution to ${\widehat{S}}$ from the local
operator ${\mathcal {O}}_S$, which we denote as ``UV contribution''
${\widehat{S}}_{\textrm{UV}}$, is given by
\begin{equation}
\displaystyle
{\widehat{S}}_{\textrm{UV}}\,\sim\,\frac{g^2}{16\pi^2}\frac{v^2}{f^2}c_S
\,,
\label{Ses0}
\end{equation}
and it is obviously incalculable within our effective theory: it can just be matched
to the observations by the choice of the input parameter $c_S$.

On top of the UV one, the ${\widehat{S}}$ parameter also receives a calculable ``IR'' contribution
${\widehat{S}}_{IR}$.
Given that the latter first emerges at the one-loop order like the divergent UV term does,
it is naively expected to be of the same order of magnitude.
If this was the case, the IR term would be completely hidden by the incalculable
one and the total ${\widehat{S}}={\widehat{S}}_{UV}+{\widehat{S}}_{IR}$ could not be
predicted. Actually, as the explicit calculation of ref.~\cite{Barbieri:2007bh} shows, there
is one IR effect which is enhanced with respect to ${\widehat{S}}_{UV}$.
This effect is an \emph{additive running} of the $c_S$ coupling from the scale
$\Lambda=4\pi f$ where it is generated with its NDA size, down to the Higgs mass $m_H$. This running
effect results in
\begin{equation}
\displaystyle
{\widehat{S}}_{\textrm{IR}}\,=\,
\frac1{6}\frac{g^2}{16\pi^2}\frac{v^2}{f^2}\log\frac{\Lambda}{m_H}\,,
\label{sBR}
\end{equation}
which is parametrically dominant, with respect to the UV contribution of eq.~(\ref{Ses0}),
due to the logarithmic factor. Formally, therefore, the ${\widehat{S}}$ parameter \emph{is}
calculable in this model, in spite of being divergent already at the leading order in the loop
expansion. In practice however the accidental numerical pre-factor of $1/6$ completely
compensates for the parametric logarithmic enhancement and ${\widehat{S}}_{\textrm{IR}}$
never dominates in practical situations. Moreover, and more importantly, the presence
of weakly coupled resonances unavoidably gives contributions to $c_S$ that are
larger than the ones estimated by NDA in eq.~(\ref{Ses0}) \cite{Giudice:2007fh}.
With respect to the
latter, the running term of eq.~(\ref{sBR}) is numerically even more irrelevant, to the point that
we will even be entitled to ignore it in the phenomenological study of our model, presented
in section~\ref{3S}. In this sense it is correct to say that, in spite of being calculable,
${\widehat{S}}$ cannot be reliably predicted within the $\sigma$-model.

The situation is similar for the ${\widehat{T}}$ parameter, because it is also
logarithmically divergent at one loop. This might be surprising at the first sight:
${\widehat{T}}$ is made of zero-momentum gauge field correlators, with no derivatives,
and eq.~(\ref{powcount}) predicts a \emph{quadratic} one-loop divergence for a local
operator with two gauge fields, no derivatives, and an arbitrary number of Higgses.
This leading divergence cancels because ${\widehat{T}}$ is forbidden by the custodial
${\textrm{SO}}(3)_c$ symmetry, which emerges from the ${\textrm{SO}}(4)$ unbroken
group of our coset. Due to the symmetry, ${\widehat{T}}$ is not
generated by the $\sigma$-model interactions, one needs insertions of the
${\textrm{SO}}(4)$-breaking vertices, which are the ones that involve the $B_\mu$
hypercharge field. The leading divergence then contains two extra powers
of $g'$, so that $\eta_{\textrm{min}}=1$ and the degree of divergence is reduced by two,
as previously explained.

The above argument can be made more rigorous and systematic by the method of
\emph{spurions}, which we will extensively use in the rest of the paper and that we
now illustrate in this simple example. Even if all what matters for ${\widehat{T}}$
is the ${\textrm{SO}}(4)$ subgroup, let us be more general and consider the full non-linearly
realized $\textrm{SO}(5)$.  The latter is an exact symmetry of the $\sigma$-model Lagrangian
(\ref{lagg0}) and it is only broken by the couplings with the SM gauge fields in eq.~(\ref{cder}).
We formally restore $\textrm{SO}(5)$ by introducing two spurions ${\mathcal G}$ and
${\mathcal G}'$ and rewriting the gauge field $A_\mu$ in eq.~(\ref{cder}) as
\begin{equation}
\displaystyle
A_\mu\,=\,
{\mathcal G}_\alpha W_\mu^\alpha\,+\,
{\mathcal G}' B_\mu\,=\,
{\mathcal G}_{\alpha\,A}T^A W_\mu^\alpha\,+\,
{\mathcal G}' _A T^A B_\mu
\,,
\label{eqabove}
\end{equation}
with the index ``$A$'' in the adjoint representation, so that the covariant derivative transforms
homogeneously under $\textrm{SO}(5)$. The other index, $\alpha$, forms a triplet of another
symmetry group, which we denote as ``elementary'' $\textrm{SU}(2)_L^0$ group,
defined as the one under which only the three $W_\mu^\alpha$ rotate, while the
Higgs and the $B$ field are invariant.
This group commutes with $\textrm{SO}(5)$, which instead only acts on the Higgs,
and is obviously also a symmetry of the gauge Lagrangian in eq.~(\ref{lagg1}). The SM
gauge group, under which both the $W$ and the Higgs transform simultaneously, is
given by the vector combination of  $\textrm{SU}(2)_L^0$ and the $\textrm{SU}(2)_L$
(see Appendix~\ref{gener}) subgroup of $\textrm{SO}(5)$. We also notice the existence of
a further symmetry, a $Z_2$ parity that changes sign to ${\mathcal G}'$ and $B_\mu$.

Of course, eq.~(\ref{eqabove}) is just a \emph{rewriting} of $A_\mu$, the physical
values of the spurions (which we will occasionally denote as ``spurion expectation values'',
or VEVs) are indeed
\begin{equation}
\displaystyle
{\mathcal G}_\alpha \,=\,g\,T_L^\alpha\,,\;\;\;\;\;\;\;\;\;\;\;\;\;\;\;
{\mathcal G}' \,=\,g'\,T_R^3
\,,
\label{svev}
\end{equation}
which gives back eq.~(\ref{cder}). The spurion ${\mathcal G}$ breaks the
total $\textrm{SU}(2)_L^0\times\textrm{SO}(5)$
symmetry group down to $\textrm{SO}(4)\simeq\textrm{SU}(2)_L\times \textrm{SU}(2)_R$,
the further breaking to the SM group is due to ${\mathcal G}' $.
The point is that \emph{before} setting them to their physical value, the spurions
have well-defined transformation properties under the symmetry and they must enter in
symmetry-preserving combinations with the other fields. Using the symmetry we can
classify the allowed local operators with two gauge fields in terms of the number of spurions
they contain. With no spurions insertions, the only operator is ${\mathcal L}^{\pi}$ in
eq.~(\ref{lagg0}) (with, obviously, covariant derivatives), which however does not contribute to
${\widehat{T}}$ because of the custodial symmetry. The leading contribution to ${\widehat{T}}$
 comes from operators such as
\begin{equation}
\displaystyle
\frac{c_T}{16\pi^2}f^2\left(\Sigma^t {\mathcal G}' D_\mu \Sigma \right)
\left(\Sigma^t {\mathcal G}' D^\mu \Sigma \right)
\,,
\end{equation}
and contains, as expected, two powers of the custodial-breaking coupling $g'$. The
coefficient $c_T$, using eq.~(\ref{powcount}), is expected to be of order unity.

Going back to the issue of calculability, the situation for ${\widehat{T}}$ is very similar to the
one for ${\widehat{S}}$. Being logarithmically divergent, ${\widehat{T}}$ is calculable because
it is dominated by an additive running effect analogous to the one for ${\widehat{S}}$.
Also in this case, however, this calculable IR contribution is numerically
not very relevant, the dominant one being radiative effects
due to the resonances. As for ${\widehat{S}}$, the ${\widehat{T}}$
parameter is calculable but not predictable within the $\sigma$-model, a description of the
resonances (in particular, of the \emph{fermionic}
resonances which give the dominant contribution) would be needed.

The mass $m_H$ of the composite Higgs boson is also beyond the reach of the $\sigma$-model.
The entire Higgs potential, indeed, diverges quadratically at one loop and it is not
calculable. Notice that, was not for symmetries, the degree of divergence would have been
even higher: the power counting of equation~(\ref{powcount}) predicts a \emph{quartic}
divergence for an operator with no gauge fields and
derivatives. But the Higgs is a NGB, and it \emph{shifts} under the non-linearly realized
transformations of $\textrm{SO}(5)$. The shift symmetry forbids the potential, which can therefore
only originate from its explicit breaking. In the language of spurions, this
means that the potential must contain powers of ${\mathcal G}$ or ${\mathcal G}'$ which
lower the degree of divergence from quartic to quadratic.
At the leading order, $g^2$ or ${g'}^2$, it is easy to classify the operators:
they are only two and read
\begin{equation}
\displaystyle
c_g\,f^4\Sigma^t {\mathcal G}_\alpha{\mathcal G}_\alpha\Sigma\,=\,\frac34
c_g\,f^4
\,g^2
\sin^2(\langle \Pi_4 \rangle/f)\;\;\;{\textrm{and}}\;\;\;\;\;
c_{g'}\,f^4\Sigma^t{\mathcal G}'{\mathcal G}'\Sigma\,=\,\frac14 c_{g'}\,f^4\,{g'}^2
\sin^2(\langle \Pi_4 \rangle/f)\,.
\label{pot0}
\end{equation}

The method of spurions is not only useful to count the powers of the gauge couplings
and to determine the NDA size of the operators, it often makes manifest more subtle
implications of the underlying symmetry. In the case at hand, the spurion analysis
has fixed the functional form of the Higgs potential, up to unknown $c_g$ and $c_{g'}$
coefficients. Unfortunately, the latter diverge quadratically, and therefore are not calculable
within the $\sigma$-model. Very much like $c_S$ and $c_T$, they should be fixed by observations
or matched to the predictions of a more complete UV theory.

\subsection{Two and three sites}\label{twothree}

We have seen that neither the Higgs potential, nor the ${\widehat{S}}$ and ${\widehat{T}}$
EWPT parameters are predictable within the $\sigma$-model. In the present section we
introduce the three-site DCHM in which, as anticipated in the Introduction, all these
observables are on the contrary perfectly calculable. The DCHM, on top of the NGB
Higgs, also describes some of the strong sector resonances.

\subsubsection{Two sites}
\label{2sit}

Before moving to the three-site model, however, let us discuss the simpler case of two sites,
in which most of the ingredients of the DCHM are already present and more
easily illustrated.
The starting point is again a non-linear $\sigma$-model, but based on a coset different from
the $\textrm{SO}(5)/\textrm{SO}(4)$ used in the previous section. We add a second
$\textrm{SO}(5)$ and consider the chiral group
$\textrm{SO}(5)_L\times\textrm{SO}(5)_R$ spontaneously broken to the vector subgroup
${\textrm{SO}(5)_V}$. This coset, $\textrm{SO}(5)_L\times\textrm{SO}(5)_R/\textrm{SO}(5)_V$,
is parametrized by an $\textrm{SO}(5)$ Goldstone matrix
\begin{equation}
\displaystyle
U[\Pi] \, =\, \exp\left[i \frac{\sqrt{2}}{f}\Pi_A T^A \right]\,,
\label{gomat}
\end{equation}
which transforms linearly under $\textrm{SO}(5)_L\times\textrm{SO}(5)_R$
\begin{equation}
\displaystyle
U[\Pi]\, \rightarrow\, U[\Pi']\,=\,\gamma_{L} U[\Pi] \gamma_{R}^t\,.
\label{trr}
\end{equation}
The SM group is embedded in the unbroken ${\textrm{SO}(5)_V}$, under
which the ten Goldstones $\Pi_A$ transform in the adjoint representation. Under
${\textrm{SO}(4)}\subset{\textrm{SO}(5)_V}$, four of these ($\Pi_{\widehat{a}}$)
form a four-plet and are identified with the Higgs field while the remaining six, in
the adjoint of $\textrm{SO}(4)$, will be removed by gauging and will not appear in
the spectrum as physical scalars.

With the threefold purpose of getting rid of these scalars, of breaking the extra
$\textrm{SO}(5)_R$ symmetry and of adding to the model a description
of the vector resonances, we gauge the $\textrm{SO}(4)$ subgroup of
$\textrm{SO}(5)_R$ by introducing six gauge fields $\widetilde{\rho}_\mu^a$.
Given that $\textrm{SO}(5)_R$ is non-linearly realized, these gauge bosons
become massive and acquire their longitudinal components by eating the
$\Pi_a$'s. Since we want to interpret the massive $\widetilde{\rho}$
states as resonances of the strongly-interacting sector, we assign them a coupling
$\widetilde{g}_*$ of the order of the typical strong sector coupling $g_\rho$.
Their mass is given by $\widetilde{m}_\rho\simeq \widetilde{g}_* f$
(see eq.~(\ref{eq:gauge masses})) and it is of
the order of the typical strong sector mass $m_\rho$.  As described in the Introduction,
$m_\rho$ is expected to be of TeV size and the coupling $g_\rho$ is
``large'', though not maximal. We also have to describe the EW bosons and,
for this purpose, we gauge the $\textrm{SU}(2)_L\times \textrm{U}(1)_Y$
subgroup of $\textrm{SO}(5)_L$ with gauge fields $W_\mu^\alpha$ and $B_\mu$.
We interpret the $W$ and $B$ fields as \emph{elementary}, their couplings $g_0$ and
$g'_0$ are much smaller than $\widetilde{g}_*$ and, as we will see below, almost
coincide with the SM $g$ and $g'$ couplings.

\begin{figure}[t]
\centering
\includegraphics[width=.4\textwidth]{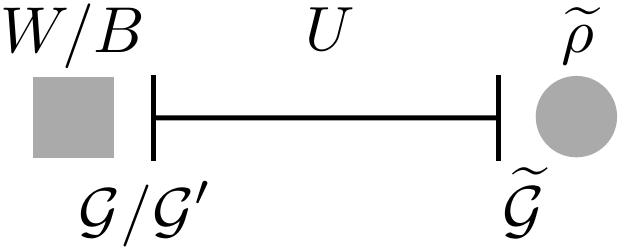}
\caption{Pictorial representation of the two-site DCHM. The Goldstone
matrix $U$ is depicted as a ``link'', {\it i.e.}~a segment with vertical lines at
the endpoints representing the global $\textrm{SO}(5)_L$ and $\textrm{SO}(5)_R$
groups. The SM and $\widetilde{\rho}$ fields are located at two different ``sites'',
represented respectively by a gray square or circle. The first site can be interpreted
as the elementary group $\textrm{SU}(2)_L^0 \times \textrm{U}(1)_Y^0$ under which only
$W$ and $B$ transform, the second one is the analogous group for $\widetilde \rho$,
$\widetilde{\textrm{SO}}(4)$. The corresponding gauge couplings, or better their associated
spurions $\mathcal{G}$, $\mathcal{G}'$ and $\widetilde{\mathcal{G}}$ are also indicated.
Their location reminds the symmetry groups under which they transform.
}\label{figstruct0}
\end{figure}

The structure of the two-site DCHM, {\it{i.e.}}~its field-content and
the gauging that we have just described, is summarized in figure~\ref{figstruct0}.
This structure corresponds to a leading order Lagrangian
$$
\displaystyle
{\mathcal L}_0\,=\,{\mathcal L}^\pi\,+{\mathcal L}_{\textrm{st}}^g\,+{\mathcal L}_{\textrm{el}}^g\,\,,
$$
where we have separated the $\sigma$-model term
\begin{equation}
\displaystyle
{\cal L}^\pi = \frac{f^2}{4}{\rm Tr}\left[(D_\mu U)^t D^\mu U\right]\,,
\label{pilag1}
\end{equation}
from the ``gauge'' part ${\mathcal L}^g={\mathcal L}_{\textrm{st}}^g+{\mathcal L}_{\textrm{el}}^g$
that only contains gauge boson fields and renormalizable interactions.
The covariant derivative, which is responsible for all the interactions among the Goldstones and
the gauge fields, is defined as
\begin{equation}
\displaystyle
D_\mu U \, = \, \partial_\mu U \,-\, i\,  A_{\mu} U\, +\,  i\, U\,\widetilde{R}_\mu\,,
\label{cd2s}
\end{equation}
with the elementary gauge field $A_{\mu}$ given, in analogy to eq.~(\ref{cder}), by
\begin{equation}
A_\mu\,=\,g_0\, W_\mu^\alpha T_L^\alpha\,+\,g'_0 B_\mu T_R^3\,,
\label{cder2sit}
\end{equation}
and with
\begin{equation}
\widetilde{R}_\mu \,=\,{\widetilde g}_*\,{\widetilde\rho}_\mu^aT_a\,.
\label{rtilda}
\end{equation}
The gauge Lagrangian of the strong sector fields is given by
\begin{equation}
\displaystyle
{\mathcal L}_{\textrm{st}}^g\,=\,-\,\frac{1}{4}{\rm Tr}\left[{\widetilde\rho}_{\mu\nu}
{\widetilde\rho}^{\mu\nu}\right]\,,
\label{glag1}
\end{equation}
with ${\widetilde\rho}_{\mu\nu} \equiv \partial_\mu {\widetilde\rho}_\nu -
\partial_\nu {\widetilde\rho}_\mu - i {\widetilde g}_*[{\widetilde\rho}_\mu, {\widetilde\rho}_\nu]$.
The Lagrangian of the elementary gauge fields, ${\mathcal L}_{\textrm{el}}^g$, is instead
given by eq.~(\ref{lagg1}).

The above equations contain mass-term mixings of the elementary $W/B$ and the composite
$\widetilde \rho$ fields. The massless combination corresponds to the gauge field of the
unbroken $\textrm{SU}(2)_L \times \textrm{U}(1)_Y$ gauge invariance and are easily
obtained by diagonalizing the mass matrix.
In this way we also find the masses of the heavy vector states:
\begin{equation}
m_L^2 = \frac{g_0^2 + {\widetilde g}_*^2}{2} f^2\,,
\qquad
m_Y^2 = \frac{{g'_0}^2 + {\widetilde g}_*^2}{2} f^2\,,
\qquad
m_C^2 = \frac{{\widetilde g}_*^2}{2} f^2\,,
\label{eq:gauge masses}
\end{equation}
which correspond respectively to the resonances associated to $T^\alpha_L$, to $T^3_R$ and to the
remaining $\textrm{SO}(5)$ generators. Furthermore, the SM gauge couplings are given by the usual expressions
\begin{equation}
\frac{1}{g^2} = \frac{1}{g_0^2} + \frac{1}{\widetilde g_*^2} \simeq \frac{1}{g_0^2}\,,
\qquad
\frac{1}{{g'}^2} = \frac{1}{{g'}_0^2} + \frac{1}{\widetilde g_*^2} \simeq \frac{1}{{g_0'}^2}\,,
\label{eq:gauge coupl 2sit}
\end{equation}
while the couplings among the resonances are of order $\widetilde g_*$.

The DCHM is a non-renormalizable effective field theory with a cut-off $\Lambda$
that can at most reach the scale $\Lambda_{\textrm{Max}}=4\pi f$, where the $\sigma$-model
interactions in eq.~(\ref{pilag1}) become non-perturbative. We assume that
$\Lambda = \Lambda_{\textrm{Max}}$ and adopt, as in the previous section, the NDA estimate
for the higher-order operators in the Lagrangian.
As shown in Appendix~\ref{NDA}, the NDA formula for
the DCHM is basically identical to eq.~(\ref{powcount}), and can be rewritten as
\begin{equation}
\displaystyle
{\mathcal L}_i\,=\,\Lambda^2\,f^2\left(\frac{\Lambda}{4\pi f}\right)^{2L}
\left(\frac{\Pi}{f}\right)^{E_\pi}
\left(\frac{gV}{\Lambda}\right)^{E_V}
\left(\frac{\partial}{\Lambda}\right)^{d}\,
\left(\frac{g f}{\Lambda}\right)^{2\eta}\,,
\label{powcount1}
\end{equation}
where ``$V$'' and ``$g$'' denote all the gauge fields ($W_\mu$, $B_\mu$ and
${\widetilde{\rho}}_\mu$) and the gauge couplings ($g_0$, $g'_0$ and
${\widetilde{g}_*}$) which are present in the theory. The dependence on
the cutoff $\Lambda$, which is left as a free parameter instead of being
set to $\Lambda_{\textrm{Max}}$, provides the degree of divergence of the
operator.

The important novelty of the two-site DCHM, with respect to the model of the previous section,
is that the Higgs boson is now a NGB with respect to \emph{two} symmetry groups instead of
one, so that its dynamics is ``doubly protected''. This mechanism is denoted in the literature as
``collective breaking''  \cite{ArkaniHamed:2001nc}. To see clearly what this implies, imagine setting
the Higgs to its VEV, which corresponds to $\langle \Pi_{\widehat{4}} \rangle \neq 0$,
with all the other components of $\Pi_A$ vanishing.
This VEV can be eliminated from the matrix $U$ in eq.~(\ref{gomat}) by
\emph{either} performing an $\textrm{SO}(5)_L$ \emph{or} an $\textrm{SO}(5)_R$ transformation,
in eq.~(\ref{trr}), along the $T^{\widehat{4}}$ generator. This implies that the dependence
on $\langle \Pi_{\widehat{4}} \rangle$
always cancels provided that \emph{any} of the two groups, or at least the subgroup generated by
$T^{\widehat{4}}$, is an exact symmetry of the theory. All the effects triggered by EWSB and
associated to the Higgs VEV, such as the ${\widehat{S}}$ and ${\widehat{T}}$ parameters and the
Higgs potential, can therefore only originate from the breaking of \emph{both} symmetries. This
leads to extra powers of the $\textrm{SO}(5)_R$-breaking coupling ${\widetilde{g}}_*$ and makes the
latter effects, following eq.~(\ref{powcount1}), further suppressed with respect to the
$\sigma$-model case of the previous section where the breaking of a single $\textrm{SO}(5)$ was sufficient. This extra
suppression reduces the degree of divergence, ${\widehat{S}}$ and ${\widehat{T}}$ therefore
become \emph{finite} at one-loop given that they were already logarithmically divergent. The potential
is instead not yet expected to be finite, because in the $\sigma$-model it was quadratically divergent.

We have seen that the broken symmetries play a crucial role in the DCHM.
In order to best exploit their implications
we introduce spurions
${\mathcal G}$, ${\mathcal G}'$ and $\widetilde{\mathcal G}$ associated to, respectively,
the gauging of the $\textrm{SU}(2)_L$ and $\textrm{U}(1)_Y$ subgroup of
$\textrm{SO}(5)_L$ and to the one of the $\textrm{SO}(4)$ subgroup of $\textrm{SO}(5)_R$.
The first ones arise, as in the previous section, from rewriting
\begin{equation}
\displaystyle
A_\mu\,=\,
{\mathcal G}_\alpha W_\mu^\alpha\,+\,
{\mathcal G}' B_\mu\,=\,
{\mathcal G}_{\alpha\,A_L}T^{A_L} W_\mu^\alpha\,+\,
{\mathcal G}' _{A_L} T^{A_L} B_\mu
\,,
\label{spurSM}
\end{equation}
with ``${A_L}$'' in the adjoint representation of $\textrm{SO}(5)_L$ and $\alpha$
in the triplet of the elementary $\textrm{SU}(2)_L^0$ group. The physical value
of these spurions is, similarly to eq.~(\ref{svev}) in the previous section
\begin{equation}
\displaystyle
{\mathcal G}_\alpha \,=\,g_0\,T_L^\alpha\,,\;\;\;\;\;\;\;\;\;\;\;\;\;\;\;
{\mathcal G}' \,=\,g'_0\,T_R^3
\,.
\label{svev2sit}
\end{equation}
The ``new'' spurion $\widetilde{\mathcal G}$ is introduced by replacing, in
equation~(\ref{cd2s}), $\widetilde{R}_\mu$ with
\begin{equation}
\widetilde{R}_\mu\,=\,\widetilde{\mathcal G}_a\widetilde{\rho}_\mu^a
\,=\,\widetilde{\mathcal G}_{A_R\,a}T^{A_R}\widetilde{\rho}_\mu^a\,,
\label{spurtilde}
\end{equation}
where ``$A_R$'' is in the adjoint of $\textrm{SO}(5)_R$. The other index, ``$a$'' is
in the adjoint of the group $\widetilde{\textrm{SO}}(4)$ which we define, similarly
to $\textrm{SU}(2)_L^0$, as the group under which only the $\widetilde{\rho}_\mu^a$
fields transform, the Goldstones and the EW bosons being invariant.
The physical value of $\widetilde{\mathcal G}$ is
\begin{equation}
\widetilde{\mathcal G}_a\,=\,\widetilde g_*\,T^a\,,
\end{equation}
and it breaks $\textrm{SO}(5)_R\times\widetilde{\textrm{SO}}(4)$ to the diagonal
${\textrm{SO}}(4)$ subgroup.

As a first application of the spurion method, let us classify the local operators
that contribute to $\widehat{S}$ and $\widehat{T}$. For what concerns $\widehat{S}$,
the leading operator is
\begin{eqnarray}
\displaystyle
\frac{c_S}{(16\pi^2)^2}{\mathcal O}_{S}&&\,=\,
\frac{c_S}{(16\pi^2)^2}
{\textrm{Tr}}\left[
A_{\mu\nu}U\widetilde{\mathcal G}_a\widetilde{\mathcal G}_a U^t A^{\mu\nu}
\right]\nonumber\\&&
\,\supset
\,
\frac34
\frac{c_S}{16\pi^2}
\frac{ \widetilde{g}_*^2}{16\pi^2}
\sin^2(\langle \Pi_4 \rangle/f)\,
g_0\, g'_0\,W^3_{\mu\nu}\, B^{\mu\nu}\,,
\label{localS2s}
\end{eqnarray}
which, as expected by collective breaking, contains two powers of $ \widetilde{g}_*$.
This reduces the degree of divergence and also the estimated NDA size: in comparison with
the $\sigma$-model result of eq.~(\ref{localS}) we have one more factor of
$(\widetilde{g}_*/4\pi)^2<1$. For ${\widehat{T}}$, the spurion analysis reveals that no
operator of order $\widetilde{g}_*^2$ contributes and that the leading term is
\begin{equation}
\displaystyle
\frac{c_T}{(16\pi^2)^3}f^2
\textrm{Tr}
\left[{\mathcal G}' U \widetilde{\mathcal G}_a\widetilde{\mathcal G}_a
U^t{\mathcal G}'D_\mu U   \widetilde{\mathcal G}_b\widetilde{\mathcal G}_b(D_\mu U)^t
\right]\,,
\label{t2}
\end{equation}
which contains \emph{four} powers of $\widetilde{g}_*$. In this case, the spurionic classification
of the operators gives a stronger suppression than what expected from collective breaking. The
conclusion is unchanged, however, $\widehat{S}$ and $\widehat{T}$ are finite at one
loop and therefore predictable within the two-site DCHM.

The Higgs potential is, on the contrary, still divergent. The divergence is associated to operators like
\begin{equation}
\displaystyle
\frac{c_g}{16\pi^2}f^4
\textrm{Tr}
\left[{\mathcal G}_\alpha{\mathcal G}_\alpha
 U \widetilde{\mathcal G}_a\widetilde{\mathcal G}_a
U^t
\right]\;\;\;\textrm{and}\;\;\;\;
\frac{c_{g'}}{16\pi^2}f^4
\textrm{Tr}
\left[{\mathcal G}'{\mathcal G}'
 U \widetilde{\mathcal G}_a\widetilde{\mathcal G}_a
U^t
\right]\,,
\end{equation}
which contain two more powers of ${\widetilde g}_*$ than the $\sigma$-model ones in
eq.~(\ref{pot0}). The presence of these additional couplings, which in turn is due to the collective
breaking of the Higgs shift symmetry, reduces the degree of divergence from quadratic to
logarithmic. To further reduce the divergence and make the potential finite we need to introduce
one additional symmetry under which the Higgs is a Goldstone. This is achieved
in the three-site DCHM, as we will now discuss.

\subsubsection{Three sites}

\begin{figure}[t]
\centering
\includegraphics[width=.7\textwidth]{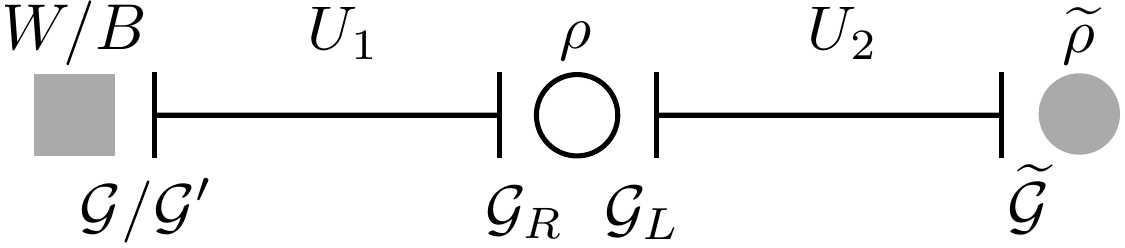}
\caption{The same as figure~\ref{figstruct0}, but for the three-site model.
}\label{figstruct1}
\end{figure}

The central ingredient for the construction of the three-site model, schematically
depicted in figure~\ref{figstruct1}, is a pair of identical $\sigma$-models,
based as before on the coset
$\textrm{SO}(5)_L\times\textrm{SO}(5)_R/\textrm{SO}(5)_V$. These are parametrized
by two $\textrm{SO}(5)$ matrices $U_1$ and $U_2$, for a total of $20$ Goldstone
bosons $\Pi^A_1$ and $\Pi^A_2$. The Goldstone Lagrangian is given, at the leading
order, by
\begin{equation}
\displaystyle
{\cal L}^\pi = \frac{f^2}{4}{\rm Tr}\left[(D_\mu U_1)^t D^\mu U_1\right]+
\frac{f^2}{4}{\rm Tr}\left[(D_\mu U_2)^t D^\mu U_2\right]\,.
\label{pilag2}
\end{equation}
The assumption that the two $\sigma$-models are identical, which led to the choice of
equal decay constants in the above equation, is equivalent to imposing a $1\leftrightarrow2$
discrete symmetry.

The symmetries of the two $\sigma$-models, \mbox{$\textrm{SO}(5)_L^1\times\textrm{SO}(5)_R^1$}
and \mbox{$\textrm{SO}(5)_L^2\times\textrm{SO}(5)_R^2$}, are broken by gauging. As in the two-site
case, the ``first'' group $\textrm{SO}(5)_L^1$ is broken by the couplings with the SM gauge bosons
and the ``last'' one, $\textrm{SO}(5)_R^2$, by the couplings with $\widetilde\rho$.
We break the remaining groups, $\textrm{SO}(5)_R^1$ and $\textrm{SO}(5)_L^2$, by gauging
their vector combination. The $10$ associated gauge fields $\rho_\mu^A$,
whose coupling is denoted by $g_*$,
become massive by eating $10$ Goldstones and are interpreted
as resonances of the strongly-interacting sector. The expressions for the masses of the
composite resonances and for the gauge couplings of the massless states will be given in
section~\ref{sec:gauge fields}.

This gauge structure, summarized in figure~\ref{figstruct1}, corresponds to the covariant derivatives
\begin{eqnarray}
&&D_\mu U_1 = \partial_\mu U_1 - i A_{\mu} U_1 + i U_1 R_{\mu}\,,\nonumber\\
&&D_\mu U_2 = \partial_\mu U_2 - i L_{\mu} U_2 + i U_2 {\widetilde{R}}_{\mu}\,,
\label{eq:covariant-derivatives}
\end{eqnarray}
where $R_\mu$ and $L_\mu$ are actually identical, $R_\mu=L_\mu=g_*\rho_\mu^AT_A$,
and $A_\mu$ is defined in eq.~(\ref{cder2sit}).
After introducing the spurions, $R_\mu$ and $L_\mu$ become different
 \begin{eqnarray}
&&R_\mu\,=\,{{\mathcal G}_R}^A{\rho}_\mu^A
\,=\,{{\mathcal G}_R}^{A_R^1\,A}\,T^{A_R^1}{\rho}_\mu^A\,,\nonumber\\
&&L_\mu\,=\,{{\mathcal G}_L}^A{\rho}_\mu^A
\,=\,{{\mathcal G}_L}^{A_L^2\,A}\,T^{A_L^2}{\rho}_\mu^A\,,
\label{spuLR}
\end{eqnarray}
and transform under different symmetries, respectively
$\textrm{SO}(5)_R^1$ and $\textrm{SO}(5)_L^2$.
The other index carried by the two spurions, ``$A$'', is associated to
transformations, which we denote as $\textrm{SO}(5)_\rho$, of the $\rho_\mu^A$ in the adjoint
with all the other fields being invariant. The spurion VEVs are
$$
{{\mathcal G}_R}^A={{\mathcal G}_L}^A=g_* T^A\,,
$$
and break $\textrm{SO}(5)_R^1\times\textrm{SO}(5)_L^2\times\textrm{SO}(5)_\rho$ to the vector combination.
As in the previous section --- eq.~(\ref{spurSM}) and eq.~(\ref{spurtilde}) --- we also
have the spurions associated to the SM couplings, ${\mathcal G}$ and ${\mathcal G}'$,
and to the one of $\widetilde{\rho}$, ${\widetilde{\mathcal G}}$.

In the three-site model the Higgs dynamics is \emph{triply} protected by the Goldstone
symmetries. Suppose indeed setting the Higgs to its VEV. This corresponds, generically,
to a constant configuration $\langle \Pi^{\widehat4}_{1,2} \rangle$ of the $\Pi^{\widehat4}_{1,2}$ Goldstone
fields. It is easy to check that this constant configuration can always be eliminated by a symmetry
transformation provided at least one of the broken symmetries is restored by setting
the corresponding coupling to zero. Let us start from $g_*$. Putting it to zero
restores $\textrm{SO}(5)_R^1$ and $\textrm{SO}(5)_L^2$, through which it is easy to
get rid of $\langle \Pi^{\widehat4}_{1} \rangle$ and $\langle \Pi^{\widehat4}_{2} \rangle$.
The situation is slightly different for ${\widetilde{g}}_*$. We have to remember that
even in the presence of the $g_*$ coupling the vector combination of
$\textrm{SO}(5)_R^1$ and $\textrm{SO}(5)_L^2$ is still unbroken, and it can be used to
set $\langle \Pi^{\widehat4}_{1} \rangle$ to zero, moving the VEV entirely on $\Pi^{\widehat4}_{2}$.
Afterwards, if ${\widetilde{g}}_*$ is set to zero, an $\textrm{SO}(5)_R^2$ transformation
can eliminate $\langle \Pi^{\widehat4}_{2} \rangle$ as well. The same holds for the elementary couplings
$g_0$ or $g'_0$. Any physical effect of the Higgs VEV, therefore, is necessarily mediated by the
three couplings $g_*$, ${\widetilde{g}}_*$ and $g_0$ (or $g'_0$).

This triple collective breaking of the Higgs shift symmetry provides a further lowering of the
degrees of divergence of all those observable which are sensitive to the Higgs VEV.
In particular, it makes the Higgs potential finite because it has to contain at least two powers
of $g_*$. Performing the spurion analysis we actually find no contributions
of ${\mathcal{O}}(g_*^2)$. The leading operators, like
\begin{equation}
\displaystyle
\frac{c_{g'}}{(16\pi^2)^3}
\textrm{Tr}\left[
{\mathcal G}'U_1{{\mathcal G}_R}^A {U_1}^t
{\mathcal G}'U_1{{\mathcal G}_R}^B {U_1}^t
\right]\cdot
\textrm{Tr}\left[
{{\mathcal G}_L}^AU_2{{\widetilde{{\mathcal G}}}}^a {U_2}^t
{{\mathcal G}_L}^B U_2{{\widetilde{{\mathcal G}}}}^a {U_2}^t
\right]
\,,
\label{pot3}
\end{equation}
contain \emph{four} powers of $g_*$. This further lowers the degree of divergence. The power
counting of eq.~(\ref{powcount1}) shows that the gauge contribution to the potential,
because of the eight powers of the
couplings, not only is finite at one loop, but it starts diverging at the three-loop order.

\subsection{Matter Sector}\label{fers}

We now have to introduce the SM fermions and couple them with the Higgs.
These states originate in our construction from \emph{elementary} degrees of
freedom, external to the strong sector, which however are coupled \emph{linearly} to some
strong sector operator.  The construction is analogous to the one of the SM vector bosons
that we described in the previous sections. The vectors arise from gauging the
SM group embedded in the strong sector ${\textrm{SO}}(4)$, which means writing down
linear couplings like $g_0\,W_\mu^\alpha J_\alpha^\mu$, where $J_\alpha^\mu$ denotes
the strong sector current operator. The linear coupling results in a \emph{mixing} of the elementary
fields with the composite resonances of the strong sector. For the vector bosons we indeed find,
in our Lagrangian, terms of the form
${\mathcal L}_{\textrm{mix}} \simeq g_0/g_\rho m_\rho^2 W_\mu^\alpha \rho_\alpha^\mu$. \footnote{Actually,
because of gauge invariance, the elementary-composite mixings arise from terms of the form
\mbox{$f^2(g_0 W^\alpha_\mu - g_* \rho^\alpha_\mu)^2$}, see for instance eq.~(\ref{pilag1}).}
Analogously, for the SM fermions we assume a mixing of the form
${\mathcal L}_{\textrm{mix}} \simeq y^f/g_\rho m_\rho \overline{f} \, \psi$, where $y^f$ is the coupling
of the elementary field with the corresponding fermionic operator and  $\psi$ generically
denotes the fermionic strong sector resonance.
Because of the mixings, the lightest state which eventually describe the SM particles are linear
combinations of the elementary states, $W$ and $f$, with the composite ones, $\rho$ and $\psi$,
realizing the so-called \emph{partial compositeness} scenario \cite{CHM1,Kaplan:1991dc}.

In the case of two sites, focusing for simplicity on the
top quark sector, we introduce only one Dirac five-plet of fermionic resonances $\widetilde\psi$.
This transforms under the $\textrm{SO}(5)_R$ $\sigma$-model group
\footnote{\label{pip1}This choice is merely conventional because by acting with the
Goldstone matrix $U$ one can convert $\widetilde\psi$ into $U\widetilde\psi$, which transforms
in $\textrm{SO}(5)_L$. Physically, $\widetilde\psi$ or $U\widetilde\psi$ equivalently describe a
${\bf5}$ of the \emph{unbroken} $\textrm{SO}(5)_V$.} and is mixed to the elementary
doublet ${q_L}=\{t_L,b_L\}$ and singlet $t_R$. The mixing term is
\begin{equation}
\displaystyle
{\mathcal{L}}_{\textrm{mix}}\,=\,y_L\,f\,{\overline{Q}_L}^I\,U_{IJ}\widetilde\psi^J\,+\,
y_R\,f\,{\overline{T}_R}^I\,U_{IJ}\widetilde\psi^J\,+\,\textrm{h.c.}\,,
\label{mixfer}
\end{equation}
where $Q_L$ and $T_R$ are the ``embeddings'' into incomplete five-plets of
$\textrm{SO}(5)_L$ of the $q_L$ and $t_R$ SM fermions.
The $\mathbf{5}$ decomposes as $\mathbf{5}=\mathbf{(2,2)}\oplus\mathbf{(1,1)}$, where
the $\mathbf{(2,2)}$ --- of $\textrm{SO}(4)$ --- consists of two $\textrm{SU}(2)_L$ doublets
of opposite $\textrm{U}(1)_R^3$ charge $T_R^3=\pm1/2$. We chose to embed $q_L$ in the
negative-charge doublet and $t_R$ in the singlet. Given the explicit form of the generators
in Appendix~\ref{gener}, the embeddings are given by
\begin{equation}
{Q_L}\,=\,\frac1{\sqrt{2}}\left(
\begin{array}{c}
b_L\\
-i\,b_L\\
t_L\\
i\,t_L\\
0
\end{array}
\right)\,,
\;\;\;\;\;
{T_R}\,=\,\left(
\begin{array}{c}
0\\
0\\
0\\
0\\
t_R
\end{array}
\right)\,.
\;\;\;\;\;
\label{embfer}
\end{equation}

If we identify the $\textrm{U}(1)_Y$ symmetry with the subgroup of $\textrm{SO}(5)_L$ generated by $T^3_R$,
as we did in the previous sections, we do not obtain the correct hypercharges for the fermions.
It is well known that this requires an extra $\textrm{U}(1)_X$ global symmetry, which acts
on the matter fields $q_L$, $t_R$ and $\widetilde \psi$ as a phase rotation with charge $X = 2/3$.
The hypercharge gauge field $B_\mu$ is now introduced by gauging the subgroup
of $\textrm{SO}(5)_L \times \textrm{U}(1)_X$ corresponding to the combination
\begin{equation}
Y = T^3_R + X\,.
\label{hych}
\end{equation}
Given that the Goldstones are not charged under the extra $\textrm{U}(1)_X$,
this change in the definition of $Y$ does not modify their couplings to $B_\mu$
and the construction discussed in the previous sections for the gauge sector is unaffected.
The $W^\alpha_\mu$ and the massive resonances $\widetilde \rho_\mu$ are included, again as before,
from gauging the $\textrm{SU}(2)_L$ subgroup of $\textrm{SO}(5)_L$ and the $\textrm{SO}(4)$
subgroup of $\textrm{SO}(5)_R$.

In order to make clear the symmetry pattern in the fermionic sector, we now
discuss how to introduce the spurions which we will then use
to analyze the fermionic contribution to the Higgs potential and to the oblique parameters
$\widehat S$ and $\widehat T$. Due to the presence of the extra $\textrm{U}(1)_X$ symmetry,
the group under which the elementary fields transform is enlarged to
$\textrm{SU}(2)_L^0 \times \textrm{U}(1)_R^0 \times \textrm{U}(1)_X^0$. The charge of the elementary fields
under $\textrm{U}(1)_R^0$ corresponds to the charge under the original subgroup
of $\textrm{SO}(5)_L$ generated by $T^3_R$, thus $q_L$ has charge $-1/2$ while $t_R$ is neutral.
The spurions are two vectors $\Delta_L$ and $\Delta_R$ in the ${\mathbf5}$ of $\textrm{SO}(5)_L$
which also transform under the elementary $\textrm{SU}(2)_L^0\times \textrm{U}(1)_R^0 \times \textrm{U}(1)_X^0$
group. The $\Delta_L$ spurion is in the $\overline{\mathbf{2}}$ with $\textrm{U}(1)_R^0$ charge $1/2$
and $X^0 = -2/3$ (the conjugate of the $q_L$ representation), while $\Delta_R$ is in the singlet
with $X^0 = -2/3$ and is neutral under $\textrm{U}(1)_R^0$.
Using these objects we can rewrite the action in eq.~(\ref{mixfer}) as
\begin{equation}
\displaystyle
{\mathcal{L}}_{\textrm{mix}}\,=\,{\overline{q}_L}^i\,\Delta^{i I}_L\,U_{IJ}\widetilde\psi^J\,+\,
\,{\overline{t}_R}\,\Delta^I_R\,U_{IJ}\widetilde\psi^J\,+\,\textrm{h.c.}\,,
\label{mixferspur}
\end{equation}
where $i$ denotes the $\textrm{SU}(2)_L^0$ index, while $I$ and $J$ are the usual $\textrm{SO}(5)$ indices.

The VEV of $\Delta_{L,R}$ breaks the strong sector group
$\textrm{SO}(5)_L\times {\textrm{U}}(1)_X$ and the elementary one
$\textrm{SU}(2)_L^0\times \textrm{U}(1)_R^0 \times \textrm{U}(1)_X^0$, preserving one
$\textrm{SU}(2)_L\times \textrm{U}(1)_Y$ subgroup that we identify with the SM group and
that we gauged by the $W$ and $B$ elementary gauge fields. The unbroken
$\textrm{SU}(2)_L$ is the vectorial combination of $\textrm{SU}(2)_L^0$ with the $\textrm{SU}(2)_L$
subgroup of $\textrm{SO}(5)_L$, while the hypercharge is the combination of the
$\textrm{U}(1)_Y^0$ subgroup of the elementary $\textrm{U}(1)_R^0 \times \textrm{U}(1)_X^0$, whose
generator is specified in eq.~(\ref{hych}), with the analogous combination
coming from $\textrm{SO}(5)_L \times \textrm{U}(1)_X$.

On top of the SM group, as already discussed, we also gauge
the ${\textrm{SO}}(4)$ subgroup of $\textrm{SO}(5)_R$ in order to introduce
the $\widetilde{\rho}$ vector resonances. This leads to the covariant derivatives
\begin{eqnarray}
&&D_\mu q_L\,=\,
\left(\partial_\mu \,-\,i\,\frac{g_0}{2}W_\mu^\alpha\sigma_\alpha \,-\,i\,\frac{g'_0}{6}B_\mu \right)q_L
\,,\nonumber\\
&&D_\mu t_R\,=\,
\left(
\partial_\mu \,-\,i\,\frac{2g'_0}{3}B_\mu\right) t_R
\,,\nonumber\\
&&D_\mu \widetilde\psi\,=\,\left(\partial_\mu \,
-\,i\,
\frac{2g'_0}{3}B_\mu\,
-\,i\,{\widetilde{\mathcal G}}_a {\widetilde{\rho}}_\mu^a \right) \widetilde\psi
\,,
\label{covderfer}
\end{eqnarray}
where the spurion ${\widetilde{\mathcal G}}$ has been introduced in the previous section.
Notice that $\widetilde\psi$ is \emph{neutral} under $\textrm{SO}(5)_L$, therefore
its covariant derivative does not contain the $W_\mu^\alpha$ gauge fields. It contains
instead a $B_\mu$ term because $\widetilde\psi$ is charged under $\textrm{U}(1)_X$
and the hypercharge is defined as in eq.~(\ref{hych}). In terms of the covariant derivatives above,
the elementary and strong sector kinetic Lagrangians read
\begin{eqnarray}
\displaystyle
&&{\mathcal L}_{\textrm{el}}^{\textrm{f}}\,=\,i\,\overline{q}_L\gamma^\mu
D_\mu q_L \,+
\,i\,\overline{t}_R\gamma^\mu
D_\mu t_R \,,\nonumber\\
&&{\mathcal L}_{\textrm{st}}^{\textrm{f}}\,=\,i\,\overline{\widetilde\psi}\gamma^\mu
D_\mu\widetilde\psi\,+\,{\widetilde{m}}^{IJ}\overline{{\widetilde\psi}}_I{\widetilde\psi}_J\,,
\label{eq:kinlag2sit}
\end{eqnarray}
where a mass-term ${\widetilde{m}}=\textrm{diag}({\widetilde{m}}_{\textrm{Q}},\,{\widetilde{m}}_{\textrm{T}})$,
different for the four-plet and the singlet components of ${\widetilde\psi}$, has also been
introduced. This mass-matrix ${\widetilde{m}}^{IJ}$ is a spurion with two indices in the fundamental
of $\textrm{SO}(5)_R$, and breaks it to its $\textrm{SO}(4)$ subgroup.

Let us pause briefly to comment on the relation of our model with the existing literature. It can be shown
by a field redefinition that the fermionic sector of the two-site DCHM is identical to the one
of ref.~\cite{Anastasiou:2009rv}. The gauge sector, on the contrary, is different because no vector
resonance ${\widetilde{\rho}}$ is taken into account in ref.~\cite{Anastasiou:2009rv}.

\begin{figure}[t]
\centering
\includegraphics[width=.4\textwidth]{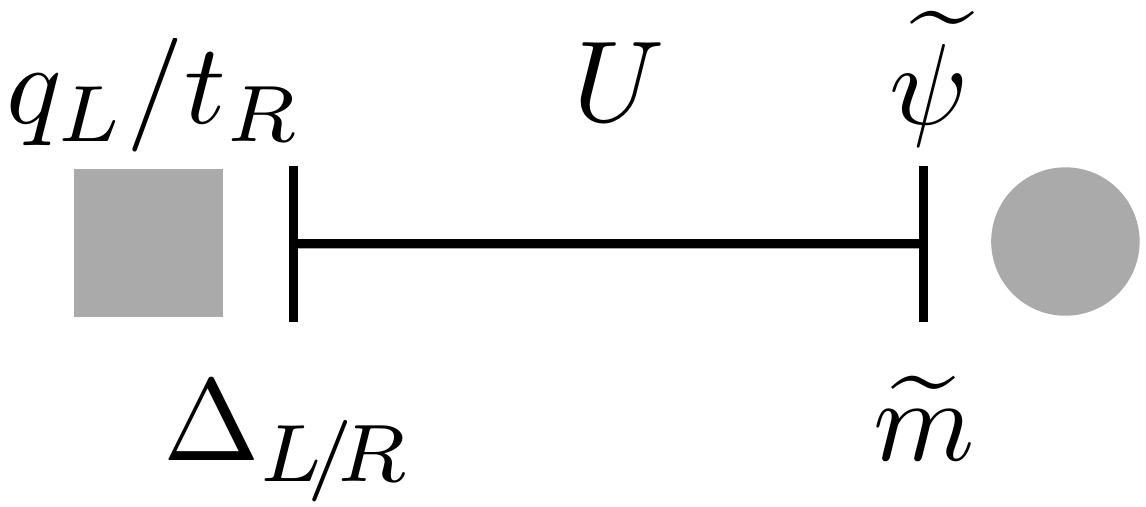}
\caption{The matter sector of the two-site DCHM. The $q_L$ and $t_R$ fermions live at
the first site, which indicates that they transform under the elementary group
$\textrm{SU}(2)_L^0\times \textrm{U}(1)_R^0\times \textrm{U}(1)_X^0$.
The resonance $\widetilde\psi$ transforms instead
with the $\textrm{SO}(5)_R$($\times\textrm{U}(1)_X$) $\sigma$-model group, and therefore it is placed at the right endpoint
of the link. The spurions $\Delta_{L/R}$ and $\widetilde m$ are also indicated, their location
reminds the symmetry groups under which they transform.
}\label{figstructfer2}
\end{figure}

The fermionic sector of the two-site DCHM, that we have just described, is summarized in
figure~\ref{figstructfer2}. With respect to the gauge sector presented in the previous section,
the only new parameters that we have introduced are $\Delta_{L,R}$ and ${\widetilde{m}}$,
and these are \emph{masses}, not new \emph{couplings}. This makes particularly easy to
generalize the NDA power-counting of eq.~(\ref{powcount1}), following the methodology
of Appendix~\ref{NDA}. The result is
\begin{equation}
\displaystyle
{\mathcal L}_i=\Lambda^2\,f^2\left(\frac{\Lambda}{4\pi f}\right)^{2L}
\left(\frac{\Pi}{f}\right)^{E_\pi}
\left(\frac{gV}{\Lambda}\right)^{E_V}
\left(\frac{\psi}{\sqrt{\Lambda} f}\right)^{E_\psi}
\left(\frac{\partial}{\Lambda}\right)^{d}
\left(\frac{g f}{\Lambda}\right)^{2\eta}
\left(\frac{\mu}{\Lambda}\right)^{\chi}\,,
\label{powcount2}
\end{equation}
where $\psi$ generically denotes the fermions $q_L$, $t_R$ or $\widetilde\psi$, while
$\mu$ is any of the masses $\Delta_{L,R}$ or ${\widetilde{m}}$. The positive integer $\chi$
counts the number of mass-term insertions and it is forced by the chiral symmetry
($\psi_L\rightarrow-\psi_L$, $\psi_R\rightarrow \psi_R$ and $\mu\rightarrow-\mu$) to
be even or odd depending on the chirality of the operator.

Now that we have introduced all the necessary tools, it is rather easy to generalize
the discussion on the calculability of the ${\widehat{S}}$
and ${\widehat{T}}$ parameters and of the Higgs potential. The newly-introduced spurions
$\Delta_{L,R}$ and ${\widetilde{m}}$ provide additional sources of breaking of the Higgs
shift symmetry, resulting in parametrically new contributions. For what concerns ${\widehat{S}}$,
a new local operator is for example
\begin{eqnarray}
\displaystyle
\frac{c_S}{(16\pi^2)^2f^2}{\mathcal O}_{S}^f&&\,=\,
\frac{c_S}{(16\pi^2)^2f^2}
{\textrm{Tr}}\left[
A_{\mu\nu}U\,\widetilde{m}^2 U^t A^{\mu\nu}
\right]\nonumber\\&&
\,\supset
\,
\frac12
\frac{c_S}{16\pi^2}
\frac{ {\widetilde{m}_{\textrm{Q}}}^2-{\widetilde{m}_{\textrm{T}}}^2}{16\pi^2f^2}
\sin^2(\langle \Pi_4 \rangle/f)\, g_0\,g'_0\,
W^3_{\mu\nu}
B_{\mu\nu}\,,
\label{localS2sfer}
\end{eqnarray}
which gets generated, with its NDA size estimated by eq.~(\ref{powcount2}), by loops
of the $\widetilde\psi$ field. Notice that the presence of two powers of the $\widetilde{m}$
spurion is due to the chiral symmetry, under which $\widetilde{m}$ is odd while
${\mathcal O}_{S}^f$ is even. Another interesting feature is the fact that
eq.~(\ref{localS2sfer}) vanishes if $\widetilde{m}_{\textrm{Q}} = \widetilde{m}_{\textrm{T}}$;
this is a consequence of the
restoration of the $\textrm{SO}(5)_R^2$ invariance in the composite fermionic sector,
which allows to remove the effects of the Higgs VEV from the Lagrangian.
Exactly like the gauge contribution in eq.~(\ref{localS2s}),
${\mathcal O}_{S}^f$ has one loop degree of divergence equal to $-2$, which means that it
is finite at one loop and starts diverging at the two-loop order. As implied by the mechanism of
collective breaking discussed in the previous section, the spurion analysis confirms that also
the fermionic contributions to the  ${\widehat{S}}$ parameter are calculable in the two-site
DCHM. For ${\widehat{T}}$, we find that the leading
contribution comes from operators like
\begin{equation}
\displaystyle
\frac{c_T}{(16\pi^2)^3}\frac1{f^4}
\left[\sum_\alpha{\Delta_L^{(\alpha)}}^\dagger
 U \,\widetilde{m}\,(D_\mu U)^t\Delta_L^{(\alpha)}
\right]^2\,,
\end{equation}
which contains \emph{four} powers of the custodial-breaking
spurion $\Delta_L$.\footnote{This can be easily proven by noticing that the spurion $\Delta_L$
has isospin $1/2$, while $\widehat T$ has isospin $2$.}
As for the gauge contribution in eq.~(\ref{t2}), the spurion analysis reveals a larger suppression of
${\widehat{T}}$ than what is implied by collective breaking, for which two powers of $\Delta_L$
would have been sufficient. Finally, again as for the gauge sector, the fermionic contributions
to the Higgs potential are still logarithmically divergent. The local operators associated to the
divergence are
\begin{equation}
\displaystyle
\frac{c_R}{16\pi^2}\Delta_R^\dagger U\,\widetilde{m}^2\,U^t\Delta_R
\;\;\;\textrm{and}\;\;\;\;
\frac{c_L}{16\pi^2}\sum_\alpha
{\Delta_L^{(\alpha)}}^\dagger U\,\widetilde{m}^2\,U^t\Delta_L^{(\alpha)}
\,,
\end{equation}
and originate, respectively, from loops of the elementary $t_R$ and $q_L$ .

\begin{figure}[t]
\centering
\includegraphics[width=.7\textwidth]{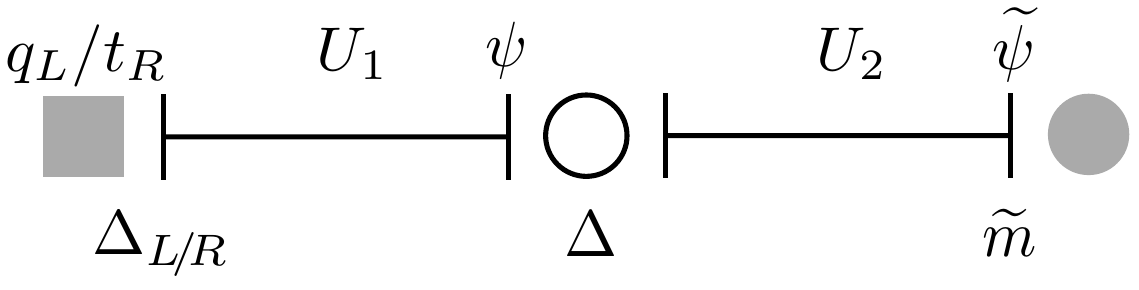}
\caption{The matter sector of the three-site DCHM.}\label{figstructfer3}
\end{figure}

To obtain a calculable Higgs potential we have to consider the three-site model.
As shown in figure~\ref{figstructfer3}, this is constructed by introducing
\emph{two} five-plets of fermionic Dirac resonances $\psi$ and $\widetilde\psi$.
The first one, $\psi$, transforms under the right group of the first link,
$\textrm{SO}(5)_R^1$, while $\widetilde\psi$ is in the fundamental of
$\textrm{SO}(5)_R^2$. The mixing Lagrangian is similar to the one of the two-site case,
with the difference that the elementary fields mix now with $\psi$ and not with $\widetilde \psi$.
Introducing the $\Delta_{L,R}$ spurions, we have
\begin{equation}
\displaystyle
{\mathcal{L}}_{\textrm{mix}}\,=\,{\overline{q}_L}^i\,\Delta^{iI}_L\,\left({U_1}\right)_{IJ}\psi^J\,+\,
{\overline{t}_R}\,\Delta_R^I\,\left({U_1}\right)_{IJ}\psi^J\,+\,
{\overline{\psi}}^I\Delta_I^{\;J}\left({U_2}\right)_{JK}{\widetilde\psi}^K
\,+\,
\textrm{h.c.}\,.
\label{mixfer3}
\end{equation}
The associated spurions, $\Delta_{L}$ and $\Delta_{R}$, transform under both
the elementary $\textrm{SU}(2)_L^0\times \textrm{U}(1)_R^0 \times \textrm{U}(1)_X^0$ and the
$\textrm{SO}(5)_L^1 \times \textrm{U}(1)_X$ group, and break the global symmetry
to the SM group as explained above in the case
of two sites. The new spurion, $\Delta$, has indices in $\textrm{SO}(5)_R^1$ and
in $\textrm{SO}(5)_L^2$. Its physical value $\Delta_I^{\;J}=\Delta\,\delta_I^{\;J}$ is
proportional to the identity, and therefore breaks
$\textrm{SO}(5)_R^1\times\textrm{SO}(5)_L^2$ to the vector subgroup.
The other terms which are present in the leading order Lagrangian are
\begin{eqnarray}
\displaystyle
{\mathcal L}_{\textrm{el}}^{\textrm{f}}\,=&& i\,\overline{q}_L\gamma^\mu
D_\mu q_L \,+
\,i\,\overline{t}_R\gamma^\mu
D_\mu t_R \,,\nonumber\\
{\mathcal L}_{\textrm{st}}^{\textrm{f}}\,=&& i\,\overline{\widetilde\psi}\gamma^\mu
D_\mu\widetilde\psi\,+\,{\widetilde{m}}^{IJ}\overline{{\widetilde\psi}}_I{\widetilde\psi}_J\nonumber\\
&& +\; i\,\overline{\psi}\gamma^\mu
D_\mu\psi\,+\,{{m}}\,\overline{{\psi}}{\psi}\,,
\label{eq:kin3sit}
\end{eqnarray}
where the covariant derivatives are defined in eq.~(\ref{covderfer}), and
\begin{equation}
D_\mu \psi\,=\,\left(\partial_\mu \,
-\,i\,
\frac{2g'_0}{3}B_\mu\,
-\,i\,
{{\mathcal G}_R}^A{\rho}_\mu^A
\right) \psi
\,.
\end{equation}
Notice that the $\psi$ mass-term $m$ does not break any symmetry, differently from
$\widetilde{m}$ which breaks  $\textrm{SO}(5)_R^2$ to its  $\textrm{SO}(4)$ subgroup.

To write the Lagrangian in eqs.~(\ref{mixfer3}) and (\ref{eq:kin3sit}) we assumed that the
strong sector is invariant under ordinary \emph{parity} ($\vec x \rightarrow -\vec x$)
and the only terms which break this
symmetry are the mixings of the elementary fields with the composite states.
If we do not impose this invariance a different mixing is allowed
between the left- and right-handed components of $\psi$ and the $\widetilde\psi$ field in
eq.~(\ref{mixfer3}).

The leading local contributions to the Higgs potential comes from operators like
\begin{equation}
\displaystyle
\frac{c_R}{(16\pi^2)^2}\frac1{f^2}\Delta_R^\dagger U_1\,\Delta\,U_2\,\widetilde{m}^2\,{U_2}^t\Delta^t
{U_1}^t\Delta_R
\,,
\end{equation}
and similarly with $\Delta_L$. These fermionic contributions are finite at one loop, and start diverging
only at two loops, differently from the gauge contributions of eq.~(\ref{pot3}) for which the
divergence was postponed to the three-loop order.

\section{Phenomenology of the three-site model}
\label{3S}

In the previous sections we established that the minimal calculable realization of the
discrete composite-Higgs scenario is the model with three sites.
This simple set-up already includes the description of two levels of resonances of
the strong sector, thus giving rise to non-trivial phenomenological aspects
which we will analyze in the present section.
To keep separate the theoretical motivation of the model and the phenomenological
analysis, we will make this section self contained and
summarize the structure of the model neglecting the technical details
that are unnecessary for the present purposes.

\begin{figure}[t]
\centering
\includegraphics[width=.7\textwidth]{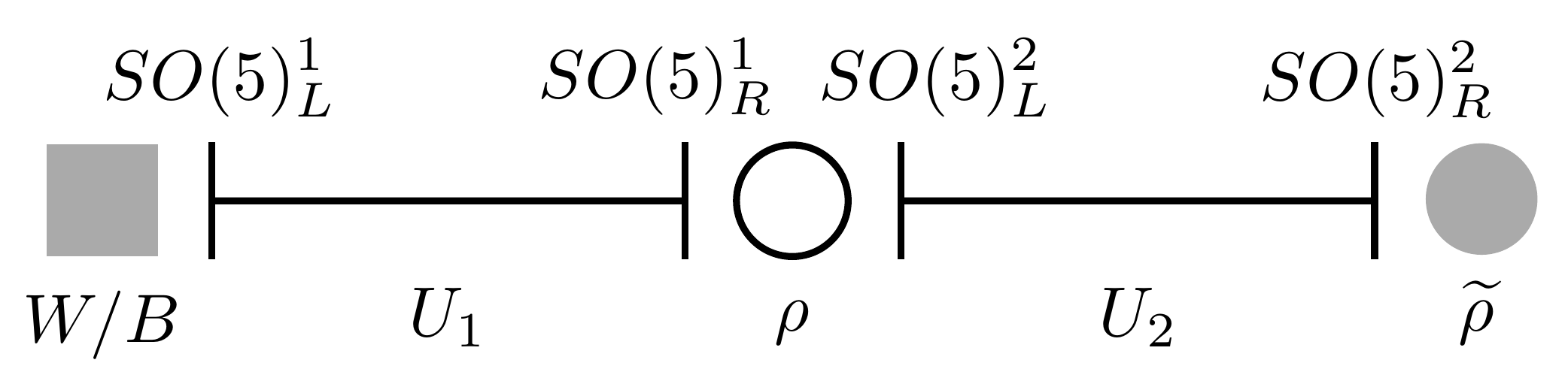}
\caption{Schematic structure of the three-site DCHM.}\label{figstruct4}
\end{figure}
The main ingredient of the three-site DCHM consists of
two replicas of the non-linear $\sigma$-model $\textrm{SO}(5)_L \times \textrm{SO}(5)_R / \textrm{SO}(5)_V$.
This symmetry structure can be connected to the usual deconstructed scenarios
by associating the global invariance to a three-site pattern,
as schematically shown in figure~\ref{figstruct4}.
Each of the two $\sigma$-models, whose Goldstone degrees of freedom are
denoted by $U_{1,2}$, is represented by a link connecting two sites. In this way we can relate
each site to a subgroup of the global invariance of the model: the first site
is associated to the $\textrm{SO}(5)_L^1$ invariance of the first $\sigma$-model, the middle
site corresponds to the $\textrm{SO}(5)_R^1 \times \textrm{SO}(5)_L^2$ subgroup and the
last site to $\textrm{SO}(5)_R^2$.
In order to accommodate the correct hypercharges for the fermionic sector,
we also need an extra $\textrm{U}(1)_X$ global factor.
This abelian factor is not associated to any of the sites and it acts on all the
fermions of the model as we will see below.
\footnote{\label{foott}From the dimensional deconstruction of a 5d composite Higgs
model we would have obtained a different structure. Analogously to the $\textrm{SO}(5)$ groups,
we would
have found \emph{four} replicas of the $\textrm{U}(1)_X$ and \emph{two} extra
Abelian $\sigma$-model $\textrm{U}(1)_L \times \textrm{U}(1)_R / \textrm{U}(1)_V$. Moreover,
two combinations of the $\textrm{U}(1)_X$'s would have been gauged and the associated
gauge field, after eating the
Abelian Goldstones, would have acquired a mass. Guided by minimality, we have chosen not
to incorporate in our model this extra structure, and therefore not to describe the additional
neutral vector resonances that would be present in the deconstructed case.}

The elementary gauge fields, as well as the vector resonances coming from the composite sector,
are introduced by gauging
suitable subgroups of the total global invariance $(\textrm{SO}(5))^4 \times \textrm{U}(1)_X$.
The elementary gauge fields $W_\mu$ and $B_\mu$ correspond to the gauging of
an $\textrm{SU}(2)_L \times \textrm{U}(1)_Y$
subgroup of $\textrm{SO}(5)_L^1 \times \textrm{U}(1)_X$,
where we identify the hypercharge generator with the combination $Y = T^{3}_R + X$.
The gauge couplings of the elementary fields are denoted by $g_0$ and $g'_0$ for the
$\textrm{SU}(2)_L$ and $\textrm{U}(1)_Y$ subgroups respectively.
Two levels of massive vector resonances are also included in the model, corresponding to a gauging at
the middle and at the last site.
At the middle site we gauge the diagonal subgroup $\textrm{SO}(5)_D$ of the global invariance
$\textrm{SO}(5)_R^1 \times \textrm{SO}(5)_L^2$, introducing the $\rho_\mu$ field
with gauge coupling $g_*$.
At the last site we add the $\widetilde\rho_\mu$
gauge fields in the adjoint of $\textrm{SO}(4)$, by gauging the corresponding
subgroup of $\textrm{SO}(5)_R^2$; its gauge coupling is denoted by $\widetilde g_*$.
The Lagrangian for the Goldstone fields of the two $\sigma$-models can be read in eqs.~(\ref{pilag2})
and (\ref{eq:covariant-derivatives}).

An equivalent form of the Lagrangian, which is more suitable
for the computation of the spectrum, of the Higgs potential and of the tree-level contribution
to the $\widehat S$ parameter, is obtained
by going to the ``holographic'' gauge. \footnote{This terminology is inspired from the extra-dimensional
holographic technique~\cite{Panico:2007qd}.}
In this gauge, the only dependence on the Goldstone degrees of freedom
appears at the first site, resulting in a particularly simple form for the Lagrangian of the
composite sector.
To reach the holographic gauge, first of all we set $U_2$ equal to the identity matrix
by using a gauge transformation of the
unbroken $\textrm{SO}(5)_D$ vector subgroup at the middle site.
All the scalar degrees of freedom are now contained in the Goldstone matrix $U_1$ of
the first $\sigma$-model, which we will denote simply with $U$ from now on.
At this point we can use the residual $\textrm{SO}(4)$ gauge invariance at the last site
to remove from $U$ the unphysical degrees of freedom and put it in the form
\begin{equation}
U = \exp\left[i \frac{\sqrt{2}}{f_\pi} h_{\hat a} T^{\hat a}\right]\,,
\label{eq:Holo_Goldstone}
\end{equation}
where the $T^{\hat a}$ generators (see Appendix~\ref{gener}) are only the ones of the
$\textrm{SO}(5)/\textrm{SO}(4)$ coset and correspond to the four Higgs components.
To reach the above gauge, while keeping the Goldstone matrix of the second
$\sigma$-model equal to the identity, we must use a combined transformation
of the diagonal subgroup $\textrm{SO}(5)_D$ at the middle site and of the $\textrm{SO}(4)$ subgroup
at the last site.
The normalization of the field $h_{\hat a}$ in eq.~(\ref{eq:Holo_Goldstone}) is chosen
such that its VEV $\langle h \rangle$ is given by $\langle h \rangle \simeq v = 246\,{\rm GeV}$
in the limit of small $\langle h \rangle/f_\pi$.
The Goldstone decay constant $f_\pi$, as shown below,
is connected to the decay constants of the original $\sigma$-models by the relation $f_\pi = f/\sqrt{2}$.

In the holographic gauge the only dependence of the Lagrangian on the Goldstone matrix comes from the
covariant derivative $D_\mu U$ (see eq.~(\ref{eq:covariant-derivatives})).
This quantity can be rewritten by using the identity
\begin{equation}
U^t D_\mu U = U^t \partial_\mu U - i U^t A_\mu U + i R_\mu
= - i A^{(U^t)} + i R_\mu\,,
\end{equation}
where $A_\mu^{(U^t)}$ corresponds to a gauge transformation of the elementary fields
(see eq.~(\ref{cder2sit})) given by $A_\mu^{(U^t)} = U^t (A_\mu + i \partial_\mu) U$.
With the above definitions, the Lagrangian in eq.~(\ref{pilag2}) can be written as
\begin{equation}
{\cal L}^\pi = \frac{f^2}{4}{\rm Tr}\left[(A^{(U^t)}_\mu - g_* \rho_\mu)^2\right]
+ \frac{f^2}{4}{\rm Tr}\left[(g_* \rho_\mu - \widetilde g_* \widetilde\rho_\mu)^2\right]\,.
\label{eq:Holo_Lag}
\end{equation}
The above expression is the right starting point to derive an effective ``holographic'' Lagrangian for the
Goldstone boson Higgs by integrating out the massive gauge resonances.
The calculation is particularly easy if we are only interested in the
two-derivative term.
This can be obtained by integrating out the resonances $\rho_\mu^{\hat a}$
corresponding to the generators of the $\textrm{SO}(5)/\textrm{SO}(4)$ coset.
The equations of motion at zero momentum are $g_* \rho_\mu^{\hat a} = \frac{1}{2} \left(A^{(U^t)}\right)^{\hat a}$,
substituting back into eq.~(\ref{eq:Holo_Lag}) we find
\begin{equation}
{\mathcal L}^\pi_{\textrm{holo}} = \frac{f^2}{8} {\rm Tr}\left[(\partial_\mu U)^t \partial^\mu U\right]\,.
\end{equation}
We can immediately read the Goldstone decay constant $f_\pi = f/\sqrt{2}$ and check that
the $h_{\hat a}$ fields in eq.~(\ref{eq:Holo_Goldstone}) have canonical normalization in the
holographic Lagrangian.

The fermionic sector of the model contains the elementary states corresponding to the
SM chiral fermions, which are introduced at
the first site, and two sets of composite resonances added at the other two sites.
We will only focus on the third quark generation, which is the one most directly
connected with EWSB. Incorporating the light families and the flavor structure
is beyond the scope of the present paper. These features could however be included
by proceeding analogously to the extra-dimensional case~\cite{EDflavor}.

Before discussing the elementary states in detail, let us consider the
fermions belonging to the composite sector. At the middle site (see figure~\ref{figstructfer3})
we introduce two multiplets $\psi_u$ and $\psi_d$, which transform in the
fundamental representation of the vector group $\textrm{SO}(5)_D$ and have $\textrm{U}(1)_X$ charges
$2/3$ and $-1/3$ respectively.
Two more multiplets in the fundamental representation of $\textrm{SO}(5)_R^2$ are introduced
at the last site $\widetilde \psi_u \in {\bf 5}_{2/3}$ and $\widetilde \psi_d \in {\bf 5}_{-1/3}$.
Given that, at the last site, the $\textrm{SO}(5)_R^2$ invariance is broken,
it is useful to introduce also a notation for the decomposition of the $\widetilde \psi_{u,d}$
multiplets in representations of the unbroken $\textrm{SO}(4) \simeq \textrm{SU}(2)_L \times \textrm{SU}(2)_R$ subgroup.
The fundamental representation of $\textrm{SO}(5)$ decomposes as
${\bf 5} = ({\bf 2}, {\bf 2}) \oplus ({\bf 1}, {\bf 1})$, thus
\begin{equation}
\widetilde \psi_u =
\left(
\begin{array}{c}
\widetilde Q_u\\
\widetilde T
\end{array}
\right)\,,
\qquad
\widetilde \psi_d =
\left(
\begin{array}{c}
\widetilde Q_d\\
\widetilde B
\end{array}
\right)\,,
\end{equation}
where $\widetilde Q_{u,d} \in ({\bf 2}, {\bf 2})$, while $\widetilde T$ and $\widetilde B$ are
the two singlets.
The Lagrangian for the composite states $\psi_u$ and $\widetilde \psi_u$
in the holographic gauge is given by
\begin{eqnarray}
{\cal L}_u^{\textrm{f}} &=& i\, \overline \psi_u \Dslash \psi_u + m^u \overline \psi_u \psi_u\nonumber\\
&& i\, \overline{\widetilde\psi}_u \Dslash \widetilde\psi_u
+ \widetilde m_{\textrm{Q}}^{u} \overline{\widetilde Q}_u {\widetilde Q}_u
+ \widetilde m_{\textrm{T}}^{u} \overline{\widetilde T} \widetilde T\nonumber\\
&& + \Delta^u \overline \psi_u \widetilde \psi_u + {\rm h.c.}\,.
\label{eq:lag_ferm_holo}
\end{eqnarray}
In the above expression we included a breaking of the $\textrm{SO}(5)_R^2$ group
through the mass terms for $\widetilde Q_u$ and $\widetilde T$ which preserve only
the $\textrm{SO}(4)$ subgroup.
The mixing in the last line of eq.~(\ref{eq:lag_ferm_holo}) comes
from the term $\Delta^u \overline \psi_u U_2 \widetilde \psi_u + {\rm h.c.}$,
which appears in the original non-gauge-fixed Lagrangian (see eq.~(\ref{mixfer3})).
The Lagrangian for the $\psi_d$ and $\widetilde \psi_d$ fields can be obtained
from the expression in eq.~(\ref{eq:lag_ferm_holo}) with obvious substitutions.

The elementary states at the first site are the
SM chiral states: the $q_L$ doublet in the ${\bf 2}_{1/6}$ representation of $\textrm{SU}(2)_L \times \textrm{U}(1)_Y$,
the $t_R$ singlet in the ${\bf 1}_{2/3}$ and the $b_R$ in the ${\bf 1}_{-1/3}$.
The terms which involve the elementary fermions, again in the holographic gauge, are given by
\begin{eqnarray}
{\cal L}^{\textrm{f}}_{\textrm{elem}} &=& i\, \overline q_L \Dslash q_L + i\, \overline t_R \Dslash t_R
+ i\, \overline b_R \Dslash b_R \nonumber\\
&& +\, \frac{y_L^u f}{\sqrt{2}} \left(\overline Q^u_L\right)^I U_{IJ} (\psi_{1R}^u)^J
+ y_R^u\, f \left(\overline T_R\right)^I U_{IJ} (\psi_{1L}^u)^J + {\rm h.c.} \nonumber\\
&& +\, \frac{y_L^d f}{\sqrt{2}} \left(\overline Q^d_L\right)^I U_{IJ} (\psi_{1R}^d)^J
+ y_R^d\, f \left(\overline B_R\right)^I U_{IJ} (\psi_{1L}^d)^J + {\rm h.c.}\,,\qquad
\label{eq:elem_lag_holo}
\end{eqnarray}
where $Q^u_L$ and $T_R$ correspond to the embedding of the $q_L$ and $t_R$ elementary fields
into the ${\bf 5}_{2/3}$ representation of $\textrm{SO}(5) \times \textrm{U}(1)_X$
as given in eq.~(\ref{embfer}), while $Q^d_L$ and $B_R$ correspond to the embedding of $q_L$
and $b_R$ into the ${\bf 5}_{-1/3}$ representation (see eq.~(\ref{ferem}) of Appendix~\ref{gener}).
The second and third lines of eq.~(\ref{eq:elem_lag_holo}) are mixings of the elementary states
with the composite particles. These mixings implement the scenario of partial compositeness in our set-up;
the $y_{L,R}^{u,d}$ couplings determine the degree of compositeness of the associated SM fields.

\subsection{The Higgs potential}

We have shown in the previous sections that in our model the Higgs effective potential is finite at one loop.
It can be computed by applying directly the Coleman--Weinberg formula or by first
deriving the holographic action, with a technique borrowed from the extra-dimensional
theories \cite{Panico:2007qd}. The result will not be reported here for brevity.
As usual in composite Higgs models, we find that the
fermion contributions dominate the potential and trigger EWSB by inducing a negative mass term for the Higgs.
Given that the $\psi_d$ and $\widetilde \psi_d$ multiplets
are only needed to generate the bottom quark mass, their
mixing with the elementary states is typically much smaller than the ones
of $\psi_u$ and $\widetilde \psi_u$, which are responsible for the
generation of the top mass. This implies that the leading contribution
to the Higgs potential comes from the $u$ multiplets
while the one from the $d$ fields can be safely neglected.

By expanding the Higgs potential at the leading order in the elementary mixings $y^u_{L,R} f$,
we find that it has the structure \footnote{The expansion parameter in eq.~(\ref{eq:leading-potential})
is given by $(y_{L,R}^u/g_\rho)^2$, as can be seen explicitly from the rewriting in
eq.~(\ref{eq:estimate-pot}). For typical configurations one finds that
the expansion parameter $(y_{L,R}^u/g_\rho)^2 \sim y_t/g_\rho$ is reasonably small.}
\begin{equation}
V(\langle h \rangle) = \left[c_L (y_L^u)^2 + c_R (y_R^u)^2\right] \frac{N_c}{16 \pi^2} \frac{m_\rho^4}{g_\rho^2}
\sin^2\left(\frac{\langle h \rangle}{f_\pi}\right) + {\cal O}\left((y^u_{L,R})^4\right)\,,
\label{eq:leading-potential}
\end{equation}
where $c_{L,R}$ are dimensionless functions of the microscopic parameters,
typically of order unity, and $N_c=3$ is the number of QCD colors.
The above formula follows from symmetries and could be derived without relying on the explicit calculation by
performing a spurion analysis based on the tools introduced in section~\ref{th}
(see also \cite{Mrazek:2011iu}).
As a consequence of ordinary parity invariance, which we imposed as a symmetry of the composite sector
(see section \ref{fers}), and of the $\textrm{SO}(5)$ group structure, we also find the exact relation
\begin{equation}
c_R = -4\,c_L\,,
\label{eq:pot-coefficients}
\end{equation}
which implies that the form of the Higgs potential at the leading order is practically fixed.

If the leading terms in eq.~(\ref{eq:leading-potential}) dominate the Higgs potential, the only
possible minima are the symmetry-preserving vacuum $\langle h \rangle = 0$
and the configuration with maximal breaking $\langle h \rangle/f_\pi = \pi/2$.
Both these possibilities are not acceptable, in particular
the configuration with maximal breaking would lead to massless fermionic states
in the spectrum and would be in strong conflict with the electroweak precision
measurements, as we will see later on.
In order to obtain a realistic EWSB we must have that $\langle h \rangle/f_\pi \simeq v/f_\pi$
is somewhat smaller than one, for which we need to rely on the effect of the
higher order contributions. The mechanism is simple: the functional form of the $(y^u_{L,R})^4$ terms
is different from the $\sin^2(\langle h \rangle / f_\pi)$ of the leading order ones,
and this allows to cancel only the Higgs mass term and not the quartic coupling, obtaining a small VEV.
But for this to happen, first of all, the two leading terms in
the potential in eq.~(\ref{eq:leading-potential}) must cancel to a good accuracy,
so that their residual contribution is comparable to the higher order terms.
The situation is exactly the same in the warped models, in particular in the MCHM$_5$.
Using eq.~(\ref{eq:pot-coefficients}), we see that to enforce the cancellation we need
\begin{equation}
y^u_L \simeq 2 y^u_R\,.
\label{eq:compositeness-relation}
\end{equation}
This relation implies that, in our model, the mixings of the elementary states $q_L$ and $t_R$
with the composite sector must have comparable sizes.
\begin{figure}
\centering
\includegraphics[width=.45\textwidth]{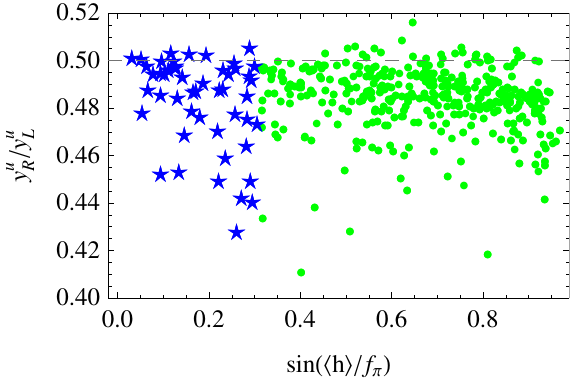}
\caption{Ratio of the mixing parameters $y_R^u$ and $y_L^u$ of the elementary
fermions $q_L$ and $t_R$ with the composite states.
The points were obtained by a scan on the microscopic parameters
with the choice $g_* = \widetilde g_* = 5$
and using the ranges $m^u, \widetilde m_{\textrm{Q}}^u, \widetilde m_{\textrm{T}}^u, \Delta^u \in [0, 8 f]$.
We further imposed a cut on the top mass $150\, \textrm{GeV} \leq m_t \leq 190\, \textrm{GeV}$.
The configurations which are
consistent with the constraints on the $\widehat S$ parameter
are highlighted by blue stars.}\label{fig:LRComp}
\end{figure}
The result in eq.~(\ref{eq:compositeness-relation}) has been checked numerically
showing that, on the points with correct EWSB,
the relation between $y^u_L$ and $y^u_R$ is verified with a {\it few}$\,\%$
accuracy in a large part of the parameter space as can be seen from fig.~\ref{fig:LRComp}.

We can get further information on the Higgs potential by
estimating the size of the $(y^u_{L,R})^4$ contributions.
Following \cite{Giudice:2007fh, Mrazek:2011iu} we have
\begin{equation}
V \simeq \frac{N_c}{16 \pi^2} m_\rho^4 \frac{y^2}{g_\rho^2} V^{(1)}(\langle h \rangle/f)
+ \frac{N_c}{16 \pi^2} m_\rho^4 \left(\frac{y^2}{g_\rho^2}\right)^2 V^{(2)}(\langle h \rangle/f)\,,
\label{eq:estimate-pot}
\end{equation}
where $y^2$ denotes generically $(y^u_L)^2$ or $(y^u_R)^2$. The first term in the equation
above, of order $y^2$, corresponds to eq.~(\ref{eq:leading-potential}).
As discussed before, the mass term coming from the $y^2$ and the $y^4$
contributions cancel between each other, but not the quartic, which then can be
estimated as \footnote{Notice that the gauge contribution to the Higgs potential,
which is parametrically smaller than the fermionic one, can sizably affect the tuning
when $\langle h \rangle/f$ is very small. The gauge contribution, however, is always
smaller than each of the $(y^u_L)^2$ and $(y^u_R)^2$ leading order fermionic contributions
to the potential in eq.~(\ref{eq:leading-potential}),
so that the cancellation mechanism which leads to eq.~(\ref{eq:compositeness-relation})
remains valid. Furthermore it is easy to see that the gauge contribution to the quartic
term in eq.~(\ref{eq:quartic_term}) is always negligible.}
\begin{equation}\label{eq:quartic_term}
V^{(4)} \sim \frac{N_c}{16 \pi^2} y^4 \langle h \rangle^4\,.
\end{equation}
From this expression we can extract the value of the Higgs mass
\begin{equation}
m_H^2 \sim 8 \frac{N_c}{16 \pi^2} y_{max}^4 \langle h \rangle^2\,,
\label{eq:app-Higgs-Mass}
\end{equation}
where we denoted by $y_{max}^4$ the maximum between $y_L^4$ and $y_R^4$.
In our case, due to eq.~(\ref{eq:compositeness-relation}), $y_L \simeq 2 y_R = y_{max}$.
On the other hand, the top mass can be estimated as usual by
\begin{equation}
m_t \simeq \frac{y^u_L y^u_R}{g_\rho} \langle h \rangle\,.
\label{eq:mtop estimate}
\end{equation}
Comparing the above equation with the Higgs mass
in eq.~(\ref{eq:app-Higgs-Mass}) we find
\begin{equation}
m_H \sim 4 \sqrt{2 N_c} \left(\frac{g_\rho}{4 \pi}\right) m_t\,.
\label{eq:mH-mt}
\end{equation}
The result is that, for typical values of the composite sector couplings,
the Higgs is relatively heavy, usually above the top mass.
\begin{figure}
\centering
\includegraphics[width=.45\textwidth]{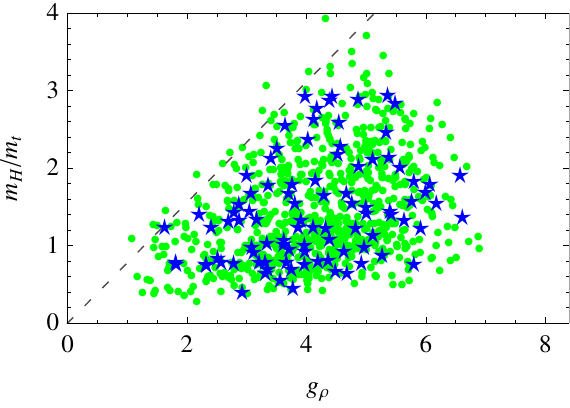}
\hspace{1.5em}
\includegraphics[width=.45\textwidth]{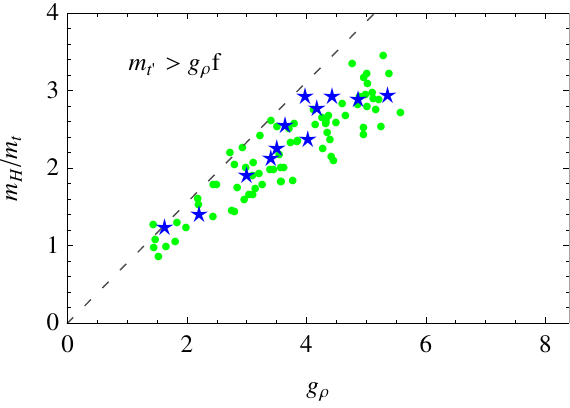}
\caption{Ratio of the Higgs an top masses as a function of the effective
coupling of the composite sector resonances. The $g_\rho$ coupling
has been estimated by the ratio of geometric average of the composite sector masses
($m^u$, $\widetilde m_{\textrm{Q}}^u$, $\widetilde m_{\textrm{T}}^u$ and $\Delta^u$)
and the Goldstone decay constant $f$. The dashed line corresponds to the relation
in eq.~(\ref{eq:mH-mt}). The plot on the right shows only the points without
light fermionic resonances (obtained by imposing the constraint $m_{t'} > g_\rho f$).
The top mass has been selected in the interval $(100 - 200)\, {\rm GeV}$.
For the range of parameters used in the scan and for the meaning of the symbols used
see the caption of fig.~\ref{fig:LRComp}.}\label{fig:mHmt}
\end{figure}
A scan on the parameter space of the model shows that the ratio $m_H / m_t$
follows an approximately linear growth in $g_\rho$ in fair agreement with the estimate
in eq.~(\ref{eq:mH-mt}), but only if no light fermionic resonance is present
(see figure~\ref{fig:mHmt}). If on the
contrary light states arise from the composite sector, eq.~(\ref{eq:mH-mt}) is usually
violated and smaller Higgs masses are obtained. Notice that it is not surprising that
our estimates are violated in this case because we assumed a common size
$m_\rho$ for all the strong sector particle masses.

\subsection{The electroweak precision measurements}\label{sec:3S EWPT}

One of the most stringent bounds on the parameter space of the
model is expected to come from the $\widehat S$ parameter, which
originates, after EWSB, from the tree-level mixing of the new strong resonances
with the SM gauge fields.
This contribution is easily computed and gives a positive
shift to $\widehat S$, whose analytic expression is
\begin{equation}
\widehat S = g_0^2 \left(\frac{3}{8 g_*^2}
+ \frac{1}{2 \widetilde g_*^2}\right) \sin^2\left(\frac{\langle h \rangle}{f_\pi}\right)\,.
\label{eq:Shat}
\end{equation}
Given that we are not performing a complete analysis of the EWPT,
we will only consider as a rough estimate the bound $\widehat S \lesssim 1.5\times 10^{-3}$.
Assuming that the gauge couplings in the composite sector are all of the same order
$g_* \sim \widetilde g_* \sim g_\rho$,
we can extract an upper bound for the value of $\sin(\langle h \rangle/f_\pi)$:
\begin{equation}
\sin\left(\frac{\langle h \rangle}{f_\pi}\right) \lesssim 0.8\, \frac{g_\rho}{4 \pi}\,.
\end{equation}
The bound on the $\widehat S$ parameter can be also directly translated in a bound on the
masses of the gauge resonances as we will discuss in subsection~\ref{sec:gauge fields}
(see eq.~(\ref{eq:S_gauge_mass})).

The other oblique EW parameter which can put constraints on the model
is $\widehat T$. Thanks to the $\textrm{SU}(2)_L \times \textrm{SU}(2)_R$ custodial symmetry,
which is left unbroken in the composite sector, the $\widehat T$ parameter vanishes
at tree-level. Its first corrections arise at one-loop level due to fermion and gauge loops.
The size of these contributions can be estimated to be of order~\cite{Giudice:2007fh}
\begin{equation}
\widehat T \sim \frac{N_c}{16 \pi^2} \frac{y_L^4}{g_\rho} \left(\frac{v}{f_\pi}\right)^2\,.
\end{equation}
In our model, using eqs.~(\ref{eq:compositeness-relation})
and (\ref{eq:mtop estimate}), we find that the mixing of the elementary fermions satisfies
the relation $y^u_L \sim y^u_R \sim \sqrt{y_t g_\rho}$, where $y_t$ is the top Yukawa coupling.
Thus we have
\begin{equation}
\widehat T \sim \frac{N_c}{16 \pi^2} y_t^2 \left(\frac{v}{f_\pi}\right)^2 \simeq 0.02 \left(\frac{v}{f_\pi}\right)^2\,.
\end{equation}
The experimental constraints on the $\widehat T$ parameter imply
a bound on $v/f_\pi$ which is comparable with the one found from $\widehat S$.
It has been shown that, in similar models (see for instance \cite{Anastasiou:2009rv}),
the contributions to $\widehat T$ from the composite fermion loops can be positive
and in many cases can help to ameliorate the compatibility with the
EW precision measurements.
The above estimate shows that $\widehat T$ is sizable and should be included in
a complete analysis of the EW constraints,
a detailed study is however beyond the scope of the present paper and we will leave it
for a future analysis.

Another possibly relevant constraint comes from the corrections to the
$Zb_L\overline b_L$ coupling. At tree-level the the deviation of this coupling
comes from two effects, the first one is the mixing at zero momentum of the $b_L$ quark with the heavy
fermionic resonances induced after EWSB and the second is
the distortion of the vertex at non-zero momentum due to higher order terms
in the effective Lagrangian.

The first effect can be described by dimension-$6$ operators
of the form $(H^\dagger D_\mu H) \overline q_L \gamma^\mu q_L$,
whose correction to the $Zb_L\overline b_L$ vertex can be estimated to be
\begin{equation}
\frac{\delta g_{Zbb}}{g_{Zbb}} \sim \frac{(y_L)^2}
{g_\rho^2}\left(\frac{v}{f_\pi}\right)^2\,,
\end{equation}
where $y_L$ denotes a generic mixing of the $q_L$ doublet with the composite fermions.
Without any suppression mechanism, for typical values of the parameters the above
correction would put strong bound on $v/f_\pi$. As explained in \cite{Agashe:2006at}, a left-right $Z_2$
symmetry $P_{LR}$ ensures that the tree-level corrections to the $Zb_L\overline b_L$ vertex
at zero momentum are absent provided that the $b_L$ field is embedded in a component
of an $\textrm{SO}(5)$ multiplet with $T_L = T_R = -1/2$.
In our model, this is the case for the $q_L$ component which lives in the
$\psi_u$ multiplet, whereas it is not true for the component in the $\psi_d$ multiplet.
The corrections can thus be estimated as
\begin{equation}\label{zbb_psi_d}
\frac{\delta g_{Zbb}}{g_{Zbb}} \sim \frac{(y^d_L)^2}{g_\rho^2}\left(\frac{v}{f_\pi}\right)^2\,.
\end{equation}
Since the $\psi_d$ multiplet is only introduced to give mass to the $b$ field,
as previously explained, the value of $y^d_L$ is typically small,
resulting in a suppressed contribution to $Zb_L\overline b_L$. In the case in which
$y^d_L \sim y^d_R$, obtaining the correct $b$ mass requires $(y^d_L)^2 \sim y_b g_\rho$, hence
\begin{equation}
\frac{\delta g_{Zbb}}{g_{Zbb}} \sim 2 \cdot 10^{-3} \left(\frac{4\pi}{g_\rho}\right)
\left(\frac{v}{f_\pi}\right)^2\,.
\end{equation}

The second tree-level correction to the $Zb_L\overline b_L$ coupling is induced at non-zero
momentum by operators of the form $\partial_\mu F^{\mu\nu} \overline q_L \gamma^\mu q_L$.
The estimate of this correction at momentum equal to the $Z$ mass is
\begin{equation}
\frac{\delta g_{Zbb}}{g_{Zbb}} \sim \frac{(y_L)^2}{g_\rho^4 f_\pi^2}m_Z^2
\sim \left(\frac{g_0^2}{g_\rho^2}\right) \frac{(y_L)^2}{g_\rho^2}\left(\frac{v}{f_\pi}\right)^2\,.
\end{equation}
In this case the contribution coming from the $\psi^u$ multiplet is not forbidden
and gives the largest correction. Using the relation
$y^u_L \sim y^u_R \sim \sqrt{y_t g_\rho}$, we get
\begin{equation}
\frac{\delta g_{Zbb}}{g_{Zbb}} \sim 2 \cdot 10^{-4} \left(\frac{4 \pi}{g_\rho}\right)^3
\left(\frac{v}{f_\pi}\right)^2\,.
\end{equation}

These results show that the constraints coming from the tree-level corrections
to the $Zb_L\overline b_L$ vertex are usually comparable or less severe than the
ones coming from the $\widehat S$ parameter.
The above estimates for the $Zb_L\overline b_L$ deviation have also been verified
by a numerical analysis.

In addition to the tree-level distortions, the $Zb_L\overline b_L$ coupling is also modified by
one-loop effects. These are potentially relevant because they
give corrections controlled by $y^u_{L,R}$ which are not protected by the
$P_{LR}$ symmetry.
These contributions can be estimated as
\begin{equation}
\frac{\delta g^{1{\textrm{-loop}}}_{Zbb}}{g_{Zbb}} \sim \frac{(y^u)^2}{16 \pi^2} \frac{(y^u_L)^2}{g_\rho^2}
\left(\frac{v}{f_\pi}\right)^2\,,
\end{equation}
where the first $(y^u)^2$ factor comes from a massive resonance circulating in the loop,
and we used $y^u_L \sim y^u_R \equiv y^u$.
Using the estimate $y^u \sim \sqrt{y_t g_\rho}$, we get
\begin{equation}
\frac{\delta g_{Zbb}^{1{\textrm{-loop}}}}{g_{Zbb}} \sim 6 \cdot 10^{-3} \left(\frac{v}{f_\pi}\right)^2\,.
\end{equation}
The 1-loop correction is therefore sizable and can be the most
relevant contribution to the deviation of the $Zb_L\overline b_L$ vertex.
The numerical analysis performed in similar models (see for example \cite{Anastasiou:2009rv})
show results in agreement with our estimates and confirms that
the constraints on $v/f_\pi$ coming from the $Zb_L\overline b_L$ coupling deviation are comparable with
the ones coming from the $\widehat S$ parameter.

\subsection{The mass spectrum: gauge fields}\label{sec:gauge fields}

The presence of new strong-sector resonances, provided they are light enough,
is the simplest and most accessible manifestation of the composite Higgs scenario at the LHC.
It is then important to discuss the properties
of the spectrum of the massive states which are present in our model.
This analysis is the subject of the present and of the next subsection.

To study the spectrum of the resonances it is useful to consider
the limit in which the couplings of the elementary fields to the composite ones
vanish. This approximation is very good in the gauge sector, given that we
assume $g_0, g'_0 \ll g_*, \widetilde g_*$.
In this limit, given that the full global invariance at the first site is restored in the gauge
sector, the Higgs VEV can be removed from the gauge Lagrangian by an $\textrm{SO}(5)^1_L$
transformation.
As a consequence, the composite gauge sector becomes invariant under the symmetry group left unbroken at the last
site, namely $\textrm{SO}(4) \simeq \textrm{SU}(2)_L \times \textrm{SU}(2)_R$, and
the heavy resonances appear in complete representations of this group.
Once the breaking generated by the mixing to the elementary fields and by the Higgs VEV is
taken into account, a small splitting is induced among the masses of the states in each multiplet.

To count the number of states we need to analyze the pattern of gauging at each of the
``composite'' sites. At the second site the diagonal $\textrm{SO}(5)_D$ global symmetry
is gauged, giving resonances in the adjoint representation of the group.
On the other hand, at the last site only the $\textrm{SO}(4)$ subgroup is gauged.
We find that the heavy gauge spectrum includes two adjoints of $\textrm{SO}(4)$ with masses
\begin{equation}
m^2_{6\pm} = \frac{f^2}{4} \left(2 g_*^2 + \widetilde g_*^2 \pm \sqrt{4 g_*^4 + \widetilde g_*^4}\right)\,,
\label{eq:gauge_mass_6}
\end{equation}
which correspond to the unbroken $\textrm{SO}(5)$ generators, and a $4$-plet in the fundamental representation
\begin{equation}
m^2_{4} = f^2 g_*^2\,,
\label{eq:gauge_mass_4}
\end{equation}
which corresponds to the broken generators.

Finally we need to consider the elementary gauge fields, which acquire mass after electroweak symmetry breaking
thanks to the mixing to the composite sector. First of all we report the expressions for the
gauge couplings corresponding to the $\textrm{SU}(2)_L \times \textrm{U}(1)_Y$ gauge symmetry
(compare with eq.~(\ref{eq:gauge coupl 2sit}))
\begin{equation}
\frac{1}{g^2} = \frac{1}{g_0^2} + \frac{1}{g_*^2} + \frac{1}{\widetilde g_*^2}\,,
\qquad
\frac{1}{{g'}^2} = \frac{1}{{g'_0}^2} + \frac{1}{g_*^2} + \frac{1}{\widetilde g_*^2}\,.
\end{equation}
By using a leading-order expansion in
the gauge couplings, $g$ and $g'$, we can compute the
masses of the $W$ and $Z$ bosons:
\begin{equation}
m_W^2 \simeq \frac{f_\pi^2}{4} g^2 \sin^2 \left(\frac{\langle h \rangle}{f_\pi}\right)\,,
\qquad
m_Z^2 \simeq \frac{f_\pi^2}{4} (g^2 + {g'}^2) \sin^2 \left(\frac{\langle h \rangle}{f_\pi}\right)\,.
\label{eq:elem_gauge_mass}
\end{equation}

The bound coming from $\widehat S$ can be immediately translated into a lower
bound on the masses of the gauge resonances. By using the expressions for
the masses of the gauge fields given in eqs.~(\ref{eq:gauge_mass_6}),
(\ref{eq:gauge_mass_4}) and (\ref{eq:elem_gauge_mass}), we can rewrite
the result given in eq.~(\ref{eq:Shat}) as
\begin{equation}\label{eq:S_gauge_mass}
\widehat S \simeq m_W^2 \left(\frac{1}{m_{6-}^2} + \frac{1}{m_{6+}^2} + \frac{1}{m_{4}^2}\right)\,.
\end{equation}
From the above expression we find the bound $m_{6\pm}, m_{4} \gtrsim 2\, {\rm TeV}$.

\subsection{The mass spectrum: fermions}\label{sec:3S ferm}

Similarly to what we did for the gauge fields, we can analyze the fermionic spectrum.
If the mixing of the elementary sector vanishes, the two sets of fields
$\{\psi_u, \widetilde \psi_u \}$ and $\{\psi_d, \widetilde \psi_d \}$
give rise to independent towers of states organized in representations
of the unbroken group $\textrm{SO}(4)$.
In the following we will only consider the fields $\psi_u$ and $\widetilde \psi_u$,
an analysis of the $d$ fields can however be obtained proceeding along the same lines.

The spectrum of the $\psi_u$ and $\widetilde \psi_u$ fields contains two $\textrm{SU}(2)_L \times \textrm{SU}(2)_R$
bidoublets with masses
\begin{equation}\label{eq:m_bid}
m^u_{bid\pm} = \frac{1}{2}\left|m^u + \widetilde m_{\textrm{Q}}^u
\pm \sqrt{(m^u - \widetilde m_{\textrm{Q}}^u)^2 + 4 (\Delta^u)^2}\right|\,,
\end{equation}
Each of these multiplets contains two massive top resonances and a bottom resonance,
moreover it includes one exotic state with electric charge $+5/3$.
The exotic state does not mix with the elementary fermions and does not couple to the Higgs,
so its mass is exactly given by eq.~(\ref{eq:m_bid}) and receives no corrections.
An interesting feature is that $m^u_{bid-}$ becomes small if the relation
\begin{equation}\label{light bidoublet}
(\Delta^u)^2 \simeq m^u \widetilde m_{\textrm{Q}}^u
\end{equation}
is satisfied. The above choice of the parameters leads to a light resonance.

The $\{\psi_u, \widetilde \psi_u\}$ tower also contains two $\textrm{SU}(2)_L \times \textrm{SU}(2)_R$ singlets
with electric charge $2/3$, whose masses are
\begin{equation}\label{m sing 2/3}
m^u_{sing\pm} = \frac{1}{2}\left|m^u + \widetilde m_{\textrm{T}}^u
\pm \sqrt{(m^u - \widetilde m_{\textrm{T}}^u)^2 + 4 (\Delta^u)^2}\right|\,.
\end{equation}
Analogously to the case of the bidoublets, the mass equation for the singlets predicts
the presence of a light state if
\begin{equation}\label{light singlet}
(\Delta^u)^2 \simeq m^u \widetilde m_{\textrm{T}}^u\,.
\end{equation}

Let us now analyze the corrections due to the mixing of the elementary states and to the EW symmetry breaking.
The elementary--composite mixings break the $\textrm{SO}(4)$ symmetry to $\textrm{SU}(2)_L$, thus the resonances are now
organized in multiplets of $\textrm{SU}(2)_L$. In particular, the bidoublet states split into two doublets
with charges $(+5/3, +2/3)$ and $(+2/3, -1/3)$. The $5/3$ states are not mixed with the elementary fields,
so their masses do not receive corrections, and, consequently, also the $(+5/3, +2/3)$ doublet has
an unchanged mass. On the other hand, the $(+2/3, -1/3)$ doublet is mixed with the elementary states
(in particular with the $q_L$ doublet) and receives a positive mass contribution proportional to $(y_L^u)^2$,
which, at leading-order in $y^u$, is
\begin{equation}
m_{1/6}^2 \simeq m_{bid\pm}^2 + \frac{(y_L^u f)^2}{4} \left(1 \pm
\frac{m^u - \widetilde m_{\textrm{Q}}^u}{\sqrt{(m^u - \widetilde m_{\textrm{Q}}^u)^2 + 4 (\Delta^u)^2}}\right)\,.
\label{eq:m16}
\end{equation}
In the same way, the singlets get a positive mass shift induced by the mixing with the $t_R$ elementary
field. The approximate expression for the shift is given by
\begin{equation}
m_{2/3}^2 \simeq m_{sing\pm}^2 + \frac{(y_R^u f)^2}{2} \left(1 \pm
\frac{m^u - \widetilde m_{\textrm{T}}^u}{\sqrt{(m^u - \widetilde m_{\textrm{T}}^u)^2 + 4 (\Delta^u)^2}}\right)\,.
\label{eq:m23}
\end{equation}

\begin{figure}
\centering
\includegraphics[width=.45\textwidth]{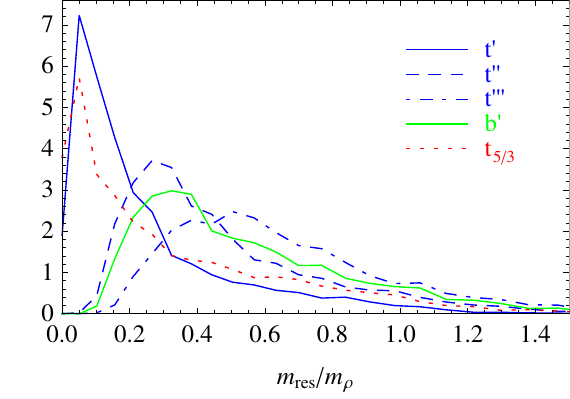}
\hspace{1.5em}
\includegraphics[width=.45\textwidth]{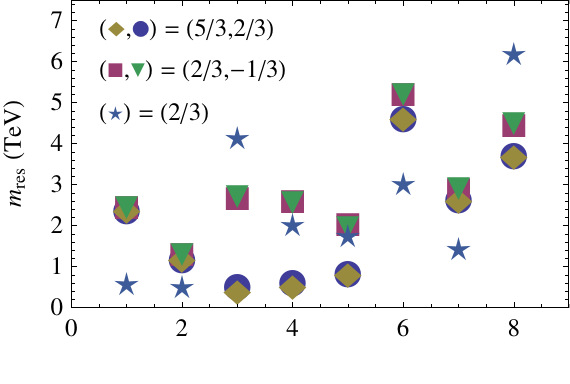}
\caption{The left panel shows the distribution of the masses of the
first level of fermionic resonances compared with the typical mass scale of the
composite sector $m_\rho$.
The mass $m_\rho$ has been estimated by geometric average of the composite sector masses
($m^u$, $\widetilde m_{\textrm{Q}}^u$, $\widetilde m_{\textrm{T}}^u$ and $\Delta^u$).
The top mass has been selected in the interval $(100 - 200)\, {\rm GeV}$.
The right panel shows the mass spectrum of the first level of fermionic
resonances for a set of sample points.
For the range of parameters used in the scan and for the meaning of the symbols used
see the caption of fig.~\ref{fig:LRComp}.}\label{fig:lightResonances}
\end{figure}

To complete the analysis we need to consider the corrections due to the breaking of the EW symmetry. Due to the
Goldstone nature of the Higgs, the effects of EWSB modify the spectrum in a very specific way.
If the mixing of the elementary fermions to the composite states vanishes, then the composite fermionic
sector has an unbroken $\textrm{SO}(5)$ symmetry, which can be used to remove the Higgs VEV from the
fermion Lagrangian. This means that the EWSB effects can modify the spectrum {\it only} through
the elementary--composite mixings $y^u_{L,R}$, so that the induced shift in the mass is of order
$(y^u_{L,R})^2 v^2$. The size of the splitting induced by the elementary mixing in eqs.~(\ref{eq:m16})
and (\ref{eq:m23}) are instead of order $(y^u_{L,R})^2 f^2$.
Given that in our model $y^u_L \sim y^u_R$, the splitting due to EWSB are all suppressed
by a factor $v^2/f^2$. This has an interesting consequence on the spectrum of the bidoublet:
its states are organized in two $\textrm{SU}(2)_L$ doublets and the splittings between
the two states in each doublet are much smaller than the mass separation between the two doublets.
This pattern can be clearly seen in the spectrum of the sample points shown in the right panel of
fig.~\ref{fig:lightResonances}. Notice that this structure is a general feature of the composite Higgs scenario
and it is due to the pNGB nature of the Higgs, therefore it is not specific of our set-up.

Another interesting property of the spectrum can be found by considering the distribution of the
masses of the resonances obtained by a scan on the parameter space of the model. As can be seen
from the left panel of fig.~\ref{fig:lightResonances}, one generically finds that the
$t'$ and the exotic state with charge $5/3$ are significantly lighter than the typical scale of the mass
parameters in the composite sector. This means that eq.~(\ref{light bidoublet}) and eq.~(\ref{light singlet})
hold accurately enough to lead to light states in a large portion of the parameter space.
The presence of these light particles, the ``top-partners'', has been noticed
already in the context of the 5d holographic models and constitutes the most
visible manifestation of the composite Higgs scenario. We have seen that
they arise in the DCHM as well.
Notice that there is typically no light $b'$ state because
its mass is lifted by corrections of order $(y^u_L)^2 f^2$.

Finally, we can derive the formula for the top mass at the leading order in the mixing
of the elementary fermions with the composite sector:
\begin{equation}\label{top mass}
m_t \simeq \frac{y^u_L y^u_R f^2}{4} \left|\frac{(\Delta^u)^2
\left(\widetilde m_{\textrm{T}}^u - \widetilde m_{\textrm{Q}}^u\right)}
{\left(m^u \widetilde m_{\textrm{Q}}^u - (\Delta^u)^2\right)
\Big(m^u \widetilde m_{\textrm{T}}^u - (\Delta^u)^2\Big)}\right|
\sin\left(\frac{2 \langle h \rangle}{f_\pi}\right)\,.
\end{equation}
The denominator in eq.~(\ref{top mass}) vanishes if one of the relations
(\ref{light bidoublet}) or (\ref{light singlet}) is satisfied. This means that, if
light states are present in the top tower, then the top mass gets an enhancement
with respect to the estimate in eq.~(\ref{eq:mtop estimate}). This effect is a consequence
of the fact that, when the top is mixed with a light resonance, its degree of compositeness
becomes larger and thus it can interact more strongly with the composite sector.
In this situation the approximate formula for the top mass is no longer accurate.

\section{Conclusions}\label{conc}

In this paper we have described the DCHM, a novel, concrete incarnation of the general
paradigm of composite Higgs. In comparison with the standard approach, which is
based on 5d gauge theories, the most obvious advantage of our model is simplicity.
The DCHM, indeed, is formulated as a standard 4d theory with a relatively small
number of fields and of free parameters. Differently from the 5d one, the DCHM is
suited for a detailed search program at the LHC because it is simple enough to be
implemented in an event generator. The DCHM also overcomes another limitation of
the 5d approach. Any 5d model hides one extra assumption, whose implications on
the phenomenology are difficult to quantify. This assumption is the choice of the metric
of the 5d space which in these constructions, that ignore gravity, is basically arbitrary.
\footnote{The 5d holographic models are typically constructed in the
truncated AdS$_5$ geometry because of the Randall-Sundrum paradigm \cite{RS}. However,
models with a different metric constitute a perfectly legitimate and phenomenologically viable
possibility \cite{CHM2flat,Becciolini:2009fu}.} In the DCHM, interpreted as the deconstructed
version of the 5d model, the ambiguity associated with the metric is encoded in the free parameters.

Notice that our model is \emph{not exactly} the three-site deconstruction of a 5d one because
it does not describe all the particles that would be present in that case. In particular, a
deconstructed model would contain massive vectors associated to the color
${\textrm{SU}}(3)_c$ (the KK gluons) and to the ${\textrm{U}}(1)_X$ local invariance in
five dimensions, as mentioned in section~\ref{3S} (see footnote~\ref{foott}).
These states are unnecessary in our model, they could however be straightforwardly included.

The DCHM is not only a simplified version of the 5d approach, it can also be regarded as an
\emph{alternative} to it. For three sites or more the DCHM has indeed the same predictive
power as the 5d one, at least for what the most relevant observables ($\widehat{S}$,
$\widehat{T}$ and the Higgs potential) are concerned. Our model offers, in this sense, a
\emph{complete} description of the composite Higgs scenario.
To further illustrate this aspect a comparison with the ``Simplified'' model of
ref.~\cite{Contino:2006nn} is probably useful. The Simplified model is a 4d description of
the first resonances and, similarly to ours, incorporates important aspects of the composite
Higgs paradigm such as the idea of partial compositeness. The main difference with
our approach is that the Higgs is \emph{not} described as a pNGB,
but instead as an ordinary scalar field. This is a serious limitation, which makes the Simplified
model fail whenever the Goldstone boson nature of the Higgs becomes important. For example,
consider the peculiar spectrum of the fermionic resonances that we encountered in section~\ref{sec:3S ferm}.
This follows from the Goldstone nature of the Higgs and it is correctly reproduced in our
set-up while it would be missed in the Simplified model. Because of these considerations
and also because the Higgs potential is not calculable, the Simplified model is not
really a complete description. The DCHM, instead, is both \emph{simple} and \emph{complete}.

Other models to which it is useful to compare our construction are the ones of ref.~\cite{Cheng:2006ht} and
\cite{Foadi:2010bu}, in the context of Little Higgs theories. Roughly speaking, the model of ref.~\cite{Cheng:2006ht} is
similar to the three-site DCHM, while the one of ref.~\cite{Foadi:2010bu} resembles the model with two
sites. There are however several important differences, which follow from the different model-building
philosophy of the DCHM. In ref.~\cite{Cheng:2006ht}, for example, only an $\textrm{SU}(2)\times\textrm{U}(1)$
 subgroup of $\textrm{SO}(5)_R^2$ is gauged at the third site, rather than an $\textrm{SO}(4)$ as
 in our case. This leaves first of all $2$ extra uneaten Goldstone bosons in the
 spectrum, leading to an extended Higgs sector. Moreover, this breaks the custodial $\textrm{SO}(4)$
 and potentially induces large tree-level contribution to the ${\widehat{T}}$ parameters. The latter
 are avoided in ref.~\cite{Cheng:2006ht} thank to T-parity, an option which however is not viable in the composite
 Higgs scenario. An even more radical difference of the DCHM is in the matter sector, where both
 ref.~\cite{Cheng:2006ht} and \cite{Foadi:2010bu} follow different (and seemingly more complicated) approaches. None of
 these approaches seems to capture the general paradigm of partial compositeness which we
 instead adopted as a crucial ingredient of the DCHM construction, as explained in sect.~\ref{fers}.

On general grounds, it is important to remind the reader that the DCHM is \emph{not} a Little Higgs
theory because no mechanism is present (or is assumed) for generating a parametrically large quartic
Higgs coupling, and therefore a parametrically small $v/f$. The DCHM only aims to describe the pNGB
Higgs and the associated resonances, the Little Higgs mechanism of generation of the quartic would
be an additional ingredient added on top of our construction.

Our work needs to the extended in two directions. One is to implement the model in an event
generator, the other is to refine our calculation of the EWPT, including the potentially relevant
radiative contributions to ${\widehat T}$ and to the $Zb_L\overline{b}_L$ vertices which we have
only estimated in section~\ref{sec:3S EWPT}. We plan to present these developments in a future publication.

\section*{Acknowledgments}

We would like to thank R.~Rattazzi, E.~Furlan and M.~Redi for useful discussions.
We also thank J.~Thaler for correspondence on his work \cite{Cheng:2006ht}.
This research is supported by the Swiss National Science Foundation under contract SNF 200020-126632.

\appendix

\section{The ${\textrm{SO}}(5)$ generators}
\label{gener}

We collect in this appendix the expressions for the $\textrm{SO}(5)$ generators in the
fundamental representation.
The generators are suitably written in a form which shows explicitly the $\textrm{SO}(4)$ subgroup:
\begin{equation}
(T^\alpha_{L,R})_{ij} = -\frac{i}{2}\left[\frac{1}{2}\varepsilon^{\alpha\beta\gamma}
\left(\delta_i^\beta \delta_j^\gamma - \delta_j^\beta \delta_i^\gamma\right) \pm
\left(\delta_i^\alpha \delta_j^4 - \delta_j^\alpha \delta_i^4\right)\right]\,,
\label{eq:SO4_gen}
\end{equation}
\begin{equation}
T^{\hat a}_{ij} = -\frac{i}{\sqrt{2}}\left(\delta_i^{\hat a} \delta_j^5 - \delta_j^{\hat a} \delta_i^5\right)\,,
\label{eq:SO5/SO4_gen}
\end{equation}
where $T^{\alpha}_{L,R}$ ($\alpha = 1,2,3$) are the
$\textrm{SO}(4) \simeq \textrm{SU}(2)_L \times \textrm{SU}(2)_R$ generators,
$T^{\hat a}$ ($\hat a = 1, \ldots, 4$) are the generators of the coset $\textrm{SO}(5)/\textrm{SO}(4)$
and the indices $i,j$ take the values $1, \ldots, 5$.
We chose to normalize the $T^{A}$'s such that ${\rm tr}[T^A, T^B] = \delta^{AB}$. With this normalization
the $\textrm{SO}(4)$ generators satisfy the commutation relations
\begin{equation}
\left[T^{\alpha}_{L,R}, T^{\beta}_{L,R}\right] = i \varepsilon^{\alpha\beta\gamma}\, T^{\gamma}_{L,R}\,,
\end{equation}
which coincide with the usual $\textrm{SU}(2)$ algebra.

The Goldstone boson matrix for the coset $\textrm{SO}(5)/\textrm{SO}(4)$ is given by
\begin{equation}
U = \exp\left[i \frac{\sqrt{2}}{f_\pi} \Pi_{\hat a} T^{\hat a}\right]
= \left(
\begin{array}{c@{\hspace{1.5em}}c}
\displaystyle I_{4 \times 4} - \frac{\vec{\Pi} \vec{\Pi}^t}{\Pi^2} \left(1 - \cos \frac{\Pi}{f_\pi}\right)
& \displaystyle \frac{\vec{\Pi}}{\Pi} \sin \frac{\Pi}{f_\pi}\\
\rule{0pt}{2em}\displaystyle - \frac{\vec{\Pi}^t}{\Pi} \sin \frac{\Pi}{f_\pi} & \displaystyle \cos \frac{\Pi}{f_\pi}
\end{array}
\right)\,,
\end{equation}
where $\Pi^2 \equiv \vec{\Pi}^t \vec{\Pi}$.

Finally, a fermion in the fundamental representation of $\textrm{SO}(5)$ can be written in the form
\begin{equation}
\psi = \frac{1}{\sqrt{2}}\left(
\begin{array}{c}
\psi_{++} + \psi_{--}\\
i \psi_{++} - i \psi_{--}\\
-\psi_{-+} + \psi_{+-}\\
i \psi_{-+} + i \psi_{+-}\\
\psi_{00}
\end{array}
\right)\,,
\label{ferem}
\end{equation}
where the $\psi_{\pm \pm}$ fields are the components of the
$\textrm{SO}(4) \simeq \textrm{SU}(2)_L \times \textrm{SU}(2)_R$ bidoublet
with charges $(\pm 1/2, \pm 1/2)$, while $\psi_{00}$ is the $\textrm{SO}(4)$ singlet.

\section{NDA Power Counting}
\label{NDA}

The NDA paradigm states that the coefficient of each operator in the Lagrangian should
be of the order of the radiative corrections it receives from the divergent diagrams,
with the divergences regulated by the physical cutoff of the effective theory.
The NDA also provides a prescription for the value of the cutoff, which should be
given by the energy scale $\Lambda_{\textrm{Max}}$
where some interaction becomes strong and invalidates the perturbative expansion.
It seems rather complicated to implement the above definition and estimate the size of the
operators because we should consider completely generic loop diagrams, constructed
with any of the infinite local vertices of our effective field theory. One can however proceed
as follows. Instead of considering all the diagrams, let us restrict to a definite
``leading order'' Lagrangian ${\mathcal L}_0$, and to loop diagrams which only involve
these leading operators. This subset of diagrams, at the price of going high enough in the
loop expansion, already contributes to all the local operators of the theory,
and can be used to estimate their NDA size. We should however worry about the effect of
the ``new'' vertices, the ones which do not appear in ${\mathcal L}_0$, and check that
inserting them into the loops does not change the estimate. Provided they are included in the
 Lagrangian with their NDA size, this is obviously the case because each insertion of a
 new vertex can be replaced by the loop diagram that generated it, and provided the estimate
 of its size. Thus the contribution from the new vertices is automatically of the
 same order of the one from ${\mathcal L}_0$ loops, and can be ignored.

To see how this works let us consider, for definiteness, the model described in
section~\ref{sigmam}, which consists of an ${\textrm SO}(5)/{\textrm SO}(4)$
$\sigma$-model coupled to the SM gauge fields. The leading order Lagrangian
is ${\mathcal L}_0={\mathcal L}^{\pi}+{\mathcal L}^{g}$, with ${\mathcal L}^{\pi}$
and ${\mathcal L}^{g}$ which are given, respectively, in eqs.~(\ref{lagg0}) and
(\ref{lagg1}).
We want to estimate the size of the one-particle irreducible ($1$PI)
diagrammatic contribution, with $L$ loops, to a generic operator with $d$
derivatives, $E_\pi$ external NGB states and $E_W$ external gauge
fields. We use a generic regulator $\Lambda$ for the UV divergences, remembering
that in order to obtain the NDA estimate $\Lambda$ must be eventually set to the physical
cutoff. In the theory under consideration, the cutoff is $\Lambda_{\textrm{Max}} = 4\pi f$,
the energy scale at which the non-linear $\sigma$-model interactions in eq.~(\ref{lagg0})
become non-perturbative.

The simplest way to obtain the result is to use dimensional analysis, counting, however,
not only the \emph{energy} dimension, but also the dimension in units of the Planck
constant $\hbar$. To see how this works, we must first of all reintroduce $\hbar$ by
replacing
$$
{\mathcal L}\;\;\;\; \rightarrow \;\;\;\; {\mathcal L}/\hbar\,,
$$
and remember that $\hbar$ acts as the loop-counting parameter. Given that
each propagator leads to one $\hbar$ and each vertex to $\hbar^{-1}$, indeed,
we have that the amputated diagrams with $E$ external states and $L$ loops
scales like
$$
\langle \phi\ldots\phi\rangle/\langle\phi\phi\rangle^E\,\propto\,\left(\hbar\right)^{L-1}\,,
$$
where $L-1=I-V$ where $I$ and $V$ are, respectively, the number of internal lines
and of vertices. What we are interested to compute are however the amplitudes of
\emph{normalized} fields
$$
\phi_N\,=\,\hbar^{-1/2}\phi\,,\;\;\;\;\;\Rightarrow \;\;\;\;\; \langle \phi_N\ldots\phi_N\rangle/\langle\phi_N\phi_N\rangle^E\,\propto\,\left(\hbar\right)^{E/2+L-1}\,.
$$
We conclude that the $1$PI diagrams we want to compute must carry an
``$\hbar$-dimension'' of $E/2+L-1$. In the basis of normalized fields, {\it{i.e.}}~after
the above rescaling, the only dependence on $\hbar$ is carried by the vertices
and it can be reabsorbed in the \emph{couplings}, which become in this way the
only $\hbar$-dimensionful objects. Imposing the $\hbar$-dimensions to match
constrains the powers of the couplings that can appear in the diagram.
When combined with the ordinary analysis of energy dimensions this
constraint will fix the
estimate of the amplitude completely.

Going back to the model of section~\ref{sigmam}, with Lagrangian
 ${\mathcal L}_0={\mathcal L}^{\pi}+{\mathcal L}^{g}$, we immediately see that,
 after the rescaling to normalized fields, all the $\hbar$ dependence can be
 reabsorbed in the decay constant $f$ and the gauge couplings $g$ and $g'$.
 The latter ones  are therefore the only couplings ({\it{i.e.}}~$\hbar$-dimensionful parameters)
 that are present in our theory and they must be combined such as to give the
 correct $\hbar$-dimension. In particular, $1/f$, $g$ and $g'$ all have the dimension of
 $\hbar^{1/2}$. Our amplitude takes the generic form
 \begin{equation}
\displaystyle
\left(\frac1{16\pi^2}\right)^L\left(\frac{\Lambda}{f}\right)^A\left(\,g\,\right)^B
\left(\frac{1}{\Lambda}\right)^{E_\pi}
\left(\frac{1}{\Lambda}\right)^{E_W}
\left(\frac{p}{\Lambda}\right)^{d}
\Lambda^4\,,
\label{nda1}
 \end{equation}
 where $p$ collectively denotes the external momenta and
 the correct energy dimension, $4-E_\pi-E_W$, has been restored with powers of
 the cutoff  $\Lambda$. Matching the $\hbar$-dimensionality requires
 \begin{equation}
 A+B\,=\,E+2L-2\,,
 \label{nda2}
 \end{equation}
 which leads to
  \begin{equation}
\displaystyle
\left(\frac{\Lambda^2}{16\pi^2f^2}\right)^L
\Lambda^2 f^2
\left(\frac{1}{f}\right)^{E_\pi}
\left(\frac{g}{\Lambda}\right)^{E_W}
\left(\frac{p}{\Lambda}\right)^{d}
\left(\frac{gf}{\Lambda}\right)^{B-E_W}
\,.
 \end{equation}

The above equation already coincides with the final result, reported in eq.~(\ref{powcount}),
it is however still to be shown that the exponent of the last term, denoted as $2\eta=B-E_W$
in eq.~(\ref{powcount}), is necessarily positive end even, as mentioned in the text. In
order to establish
this point, one needs to look more closely to where the ``$B$'' powers of $g$ could have
originated from. They could have came from the vertices in ${\mathcal L}^{\pi}$
(in eq.~(\ref{lagg0}), with the covariant derivative in eq.~(\ref{cder})), in a number which
equals the total number of $W$ lines present in the vertices
$$
B_{\pi}\,=\,\sum_{i\in \pi} E^i_W\,,
$$
or from the renormalizable gauge vertices in ${\mathcal L}^{g}$ (eq.~(\ref{lagg1})). In this
second case, their number is
$$
B_{g}\,=\,\sum_{i\in g} E^i_W-2V_g\,,
$$
where the sum runs now only on the vertices in ${\mathcal L}^{g}$, whose total
number we denote as $V_g$. Summing the two contributions, and using the relation
\begin{equation}
\sum_i E^i_W\,=\, 2 I_W + E_W\,,
\label{eq:cons of ends}
\end{equation}
where $I_W$ is the number of internal $W$ lines, we finally obtain
\begin{equation}
B\,=\,B_\pi+B_g\,=\,2\left(I_W-V_g\right)+E_W\,.
\label{Bfor}
\end{equation}
Eq.~(\ref{eq:cons of ends}) follows, as usual, from the fact that each $W$ line
either connects two vertices or one vertex and one external state.
We have now shown that $\eta=I_W-V_g$ is an integer, it is not difficult to realize
that it must be also be positive. This is because each ``g'' vertices only involves external
$W$ lines, so that in order to form part of an $1$PI diagram it must be attached to two
$W$ propagators at least. This implies $I_W\geq V_g$.

The generalization to the other models considered in the text is straightforward,
consider for instance the two-site DCHM described in section~\ref{2sit}. Also in this
case we have a set of gauge couplings ($g_0$, $g'_0$ and $\widetilde{g}$) of zero energy
dimension, plus the $\sigma$-model dimensionful coupling $1/f$. Equation~(\ref{nda1}),
with the condition~(\ref{nda2}), therefore holds true in this case as well. Also, most of the
considerations that led to eq.~(\ref{Bfor}) remain valid. We can still count the number of
gauge couplings by separating vertices from the $\sigma$-model (${\mathcal L}^{\pi}$
in eq.~(\ref{pilag1})) and from the renormalizable gauge interactions
(${\mathcal L}^{g}={\mathcal L}_{\textrm{el}}^g+{\mathcal L}_{\textrm{st}}^g$,
see eq.~(\ref{glag1})). The only difference is that extra powers of the gauge coupling can
now arise from the denominators of massive gauge field
propagators. By expanding a divergent loop integral one has indeed  the structure
$\Lambda^D+m_\rho^2\Lambda^{D-1}+\ldots$, with $m_\rho\simeq g_\rho f$.
These additional powers of $g$ invalidate eq.~(\ref{Bfor}), they are however
unavoidably positive and even, so that the conclusion that $B$ is positive and even
 remains unchanged.


\end{document}